\documentclass[fp,twocolumn]{jpsj3} 
\usepackage{latexsym,amsmath,amssymb,graphicx,bm,here,braket,subfiles,mathtools,comment}
\usepackage{txfonts}
\usepackage[dvipdfmx]{color}
\usepackage{url}

\graphicspath{{./fig_light/}}
\newcommand{\tblue}{}
\newcommand{\tred}{}

\begin{document}

\title{Superconducting Correlations in the One-Dimensional Kondo Lattice Models under Magnetic Fields}
\author{Kohei Suzuki and Kazumasa Hattori}

\inst{Department of Physics, Tokyo Metropolitan University, 1-1 Minami-osawa, Hachioji, Tokyo 192-0397, Japan}

\abst{
	We analyze superconducting correlations in the one-dimensional Kondo lattice models with Ising anisotropy 
under transverse magnetic fields, using the density matrix renormalization group.
For the spin-1/2 local spin model, the Ising anisotropy is introduced by the ferromagnetic Ising interaction between the local spins,
while for the spin-1 model, it is taken by the single-ion anisotropy.
The magnetic properties under the transverse fields for the spin-1/2 model are very similar to those for the spin-1 model [K. Suzuki and K. Hattori, J. Phys. Soc. Jpn. {\bf 88}, 024707 (2019).].
For the superconducting correlations, we analyze various Cooper pairs within nearest-neighbor pairs including composite ones between the local spins and the electrons.
We find that, for the spin-1/2 model, the superconducting correlations are highly enhanced in the Tomonaga-Luttinger liquid state near the Kondo-plateau phase,
where the conduction electrons and the local spins are strongly coupled with a finite spin gap for the Ising axis.
This is a clear contrast to the model under the longitudinal magnetic fields, where there are no noticeable superconducting correlations.
Competitions between the transverse magnetic field and the Kondo singlet formation lead to this enhanced superconducting correlations. 
\tred{
For the spin-1 model, the single-ion anisotropy suppresses the superconducting correlation and there is no noticeable enhancement. We also examine the large Kondo exchange coupling limit. For the moderate ferromagnetic Ising interaction between the local spins, we find that another type of superconducting correlation is enhanced inside the ferromagnetic phase.} We discuss a possible relation between our results and reentrant superconductivity in U-based ferromagnetic superconductors under transverse magnetic fields.
}

\maketitle

\section{Introduction}\label{sec:Introduction}
Ferromagnetic (FM) superconductivity has attracted great attention after the discovery of the first report in UGe$_2$.\cite{UGe2} 
For about two decades, both experimental and theoretical studies have been carried out intensively,
and at the present time, there are several candidate materials for the FM superconductors such as
UGe$_2$\cite{UGe2,PRB.63.144519}, URhGe\cite{URhGe}, UCoGe\cite{UCoGe}, UIr\cite{JPSJ.73.3129}, and so on.
UTe$_2$\cite{Science.365.684,JPSJ.88.043702}, recently discovered to be a superconductor, is also considered to possess similar physics to these lines of materials\cite{Rev_Exp3}.

An important aspect in these compounds is that they all show the Ising anisotropy in their magnetic responses\cite{PS75.546,PRL.100.077002,PRB.83.195107}.
Owing to this, the effects of the longitudinal and transverse magnetic fields are distinct on their superconducting (SC) properties. 
For URhGe and UCoGe, these aspects have been extensively studied in the early stage of the experiments.
For example, URhGe shows a FM transition at $T_c\sim 10$ K at zero magnetic field $H=0$ with the magnetic moment parallel to the $c$-axis,
and a SC transition occurs at $T_{\rm sc}\sim 1$ K.\cite{URhGe}
Upon applying the transverse magnetic field along the $b$-axis,
the superconductivity disappears at $H_b\sim 2$ T, while it reappears at around $H_b\sim 8$ T\cite{URhGe_RSC}.
The high-field superconductivity is related to the FM transition under the transverse fields.
This is called reentrant superconductivity and is the main subject of this paper.
Detailed NMR experiments clarify the profile of the spin fluctuation around the tricritical point of the ferromagnetism.\cite{PRL.114.216401,PRB.93.201112} Similar tendencies are also reported in UCoGe\cite{PRL.105.206403,PRL.107.187202,PRL.108.066403},
where the superconductivity continues to exist with a huge increase in the transition temperature.\cite{JPSJ.78.113709}

For theoretical sides, FM superconductivity has been intensively 
studied\cite{PhysRevB.22.3173,PhysRevB.67.024515,PhysRevB.66.134504,Rev_Theo1,Rev_Theo2},
starting from the analysis of the superconductivity in UGe$_2$\cite{PRL.90.167005,PRL.94.097003,PRB.77.184511,PRB.79.064501}.
For the reentrant superconductivity, there are several theoretical analyses, such as the Ising-magnon mechanism\cite{PRB.87.064501}, superconductivity mediated by phenomenological spin fluctuations\cite{JPCS.449.0120029,PRB.93.174512}, and the effect of a Lifshitz transition\cite{PRL.121.097001} in recent years.
However, the lack of microscopic analyses prevents us from deep understanding of the reentrant superconductivity.
It is highly desired to clarify whether SC correlations are developed or not in a standard model describing the Ising ferromagnets.

In our previous study\cite{JPSJ.88.024707}, we have clarified the magnetic phase diagram of the one-dimensional spin-1 Kondo lattice model (KLM) with the single-ion anisotropy under the transverse magnetic field.
In this study, we further extend our analysis to the SC correlations in the one-dimensional KLMs by using the density matrix renormalization group (DMRG) method\cite{PRL.69.2863,PRB.48.10345},
which is the most reliable numerical method to analyze one-dimensional correlated systems.
The DMRG method enables us to examine the enhancement of the SC correlations and to identify the possible mechanism related to the true ground states of the system.
Although the one-dimensional models are too simplified to represent the real situations,
the tendency of the SC correlations affects the bulk properties in the higher dimensions.
Thus, we believe that the asymptotically exact DMRG data in the one-dimensional models are important as a starting point of the discussions about the higher-dimensional systems.

From the theoretical point of view, there are several studies about the KLM with Ising anisotropy\cite{PRL.105.036403,Yamamoto15704,PRB.97.245119}
and the Kondo-Heisenberg model\cite{PRB.75.165110,PhysRevB.90.045125,S.R.7.11924,PRL.119.247203}.
However, the superconductivity has not been well studied in these works.
For the conventional KLMs, i.e., without the exchange interaction between the local spins,
the superconductivity is analyzed by several authors.
For example, the s-wave superconductivity is claimed in the dynamical mean-field theory (DMFT) study with the numerical renormalization group as the impurity solver\cite{Bodensiek_2010,PRL.110.146406}.
However, different superconducting states are reported by the more elaborated dual-fermion approach by Otsuki\cite{PRL.115.036404},
and he shows that an odd-frequency superconductivity in addition to the conventional one is realized near the antiferromagnetic (AFM) phase.
For the two-channel KLM, Hoshino and Kuramoto demonstrate that composite pairs,
which corresponds to the odd-frequency pairs in the conventional Cooper pair amplitude,
indeed, emerge in the infinite dimension\cite{PRL.112.167204}.
\tred{ As DMRG studies for the models related to the KLM, the KLM with an attractive Hubbard interaction is analyzed.\cite{PhysRevB.79.220513}.} 
For the extended Anderson lattice model in one-dimension, Watanabe et al., discuss possible superconductivity near the valence critical point by using the DMRG\cite{JPSJ.75.043710}. 
Recently, the superconductivity in the FM phases and its magnetic-field effect in the orbital degenerate Anderson lattice model are analyzed for discussing UGe$_{2}$.\cite{PRB.97.224519,PRB.99.205106}
Compared to the above theoretical studies, 
our interest in this paper is about the superconductivity under transverse magnetic fields in the Ising anisotropic KLM.
This kind of aspect has never been discussed in the numerical studies of the KLM and is worthwhile to be clarified.

This paper is organized as follows.
In Sect. \ref{sec:Model}, we will introduce one-dimensional KLMs with the local spin-1/2 and spin-1 degrees of freedom;
with Ising anisotropy in the exchange interactions for the former and with the single-ion anisotropy for the latter.
Sect. \ref{sec_mag} is devoted to the introduction of the basic properties of the two models under the transverse magnetic fields,
where the magnetic phase diagrams are discussed.
In Sect. \ref{sec_sc}, first, we will show the formulation of composite Cooper pairs in our DMRG study,
and then we will discuss the numerical results of the SC correlations.
In Sect. \ref{sec_dis}, the effects of the longitudinal magnetic field and the large Kondo coupling limit are examined.
We will also compare our results with the experimental data for U-based FM superconductors.
We will finally give concluding remarks in Sect. \ref{sec_sum}.

\section{Model}\label{sec:Model}

In this section, we introduce two types of one-dimensional KLMs\cite{RMP.69.809}.
The one is the spin-1/2 KLM with the Ising-type interaction under  transverse fields.
The Hamiltonian reads
\begin{align}
	\hat{H}^{S=1/2}_{\text {KLM}}&=-t\sum_{j=1}^{N-1}\sum_{\sigma=\uparrow,\downarrow}\left(\hat{c}_{j, \sigma}^{\dagger}\hat{c}_{j+1, \sigma}+\text {h.c.}\right)
	+J\sum_{j=1}^{N}\hat{\bm{s}}_{j}\cdot\hat{\bm{S}}_{j}\nonumber \\
	&-h\sum_{j=1}^{N}\left(\hat{s}^{x}_{j}+\hat{S}^{x}_{j}\right)
	-I\sum_{j=1}^{N-1}\hat{S}^{z}_{j}\hat{S}^{z}_{j+1}.
	\label{S1_HAM}
\end{align}
The other is the spin-1 KLM with the uniaxial anisotropy under transverse fields.\cite{JPSJ.88.024707} 
The Hamiltonian is 
\begin{align}
	\hat{H}^{S=1}_{\text {KLM}}&=-t\sum_{j=1}^{N-1}\sum_{\sigma=\uparrow,\downarrow}\left(\hat{c}_{j, \sigma}^{\dagger}\hat{c}_{j+1, \sigma}+\text {h.c.}\right)
	+J\sum_{j=1}^{N}\hat{\bm{s}}_{j}\cdot\hat{\bm{S}}_{j}\nonumber \\
	&-h\sum_{j=1}^{N}\left(\hat{s}^{x}_{j}+\hat{S}^{x}_{j}\right)
	-D\sum_{j=1}^{N}\left(\hat{S}^{z}_{j}\right)^{2}.
	\label{S2_HAM}
\end{align}
Here, $N$ is the system size and $\hat{c}_{j,\sigma}$ is the annihilation operator of the conduction electron at the $j$ site with the spin $\sigma = \uparrow,\downarrow$.
$\hat{\bm{s}}_{j}=(\hat{s}^{x}_{j}, \hat{s}^{y}_{j}, \hat{s}^{z}_{j})$ represents the $S=1/2$ spin operator of the conduction electron,
and $\hat{\bm{S}}_{j}=(\hat{S}^{x}_{j}, \hat{S}^{y}_{j}, \hat{S}^{z}_{j})$ is the local spin operator.
The magnitudes of the local spins are $S=1/2$ for Eq. (\ref{S1_HAM}) and $S=1$ for Eq. (\ref{S2_HAM}).
The local spin states at the site $j$ are denoted as $(\ket{\uparrow}_{j},\ket{\downarrow}_{j})$ for the spin-1/2 KLM and 
$(\ket{\Uparrow}_{j},\ket{\mathbb O}_{j},\ket{\Downarrow}_{j})$ for the spin-1 KLM.
$I > 0$ and $D > 0$ represent the Ising interaction and the uniaxial anisotropy, respectively.
$h > 0$ is the transverse magnetic field and $J > 0$ is the antiferromagnetic Kondo exchange coupling.
$t$ is the nearest-neighbor hopping and we set $t=1$ as a unit of energy.
Since the Hamiltonians Eq. (\ref{S1_HAM}) and Eq. (\ref{S2_HAM}) have a spin-inversion symmetry\cite{JPSJ.88.024707},
every state is labeled by the eigenvalues of the total electron number $N_c$ and the spin-inversion parity $P=\pm 1$.
See Appendix\ref{append} for the detail.

We note the following two differences between the spin-1/2 KLM [Eq. (\ref{S1_HAM})] and the spin-1 KLM [Eq. (\ref{S2_HAM})].
The one is the magnitude of the local spins and the other is the type of Ising interactions.
The spin-1/2 KLM has the interaction between the local spins, while the spin-1 KLM has the single-ion anisotropy.

We analyze the ground-state properties of both the KLMs with the open boundary condition, using the DMRG method.
We use the initial guess for the ground-state eigenvector in the finite DMRG algorithm\cite{PRL.77.3633} to reduce the computation time.
The local electron density, spin--spin correlation functions, and SC correlations are calculated for the system size up to $N=120$.
We keep the cutoff from $m=300$ to 700 in the truncations and the maximum truncation error is $\sim 10^{-5}$.

\section{Magnetic Properties}
\label{sec_mag}
In this section, we will show the ground-state phase diagrams 
and magnetic properties of both the KLMs.
First, we will show the $h$--$J$ ground-state phase diagrams in Sect. \ref{sub_pd}. The spin--spin correlation 
functions are also presented and they can characterize each of the ground state.
Then, we will show magnetization curves and explain ``Kondo plateau" in Sect. \ref{sub_mag}.
We will mainly discuss the results for the spin-1/2 KLM for $n_c=N_c/N=1/2$.
For other fillings, the results combined with the SC correlations are summarized in Sect. \ref{sec_pow}.
Thus, we will omit to show them in this section.
Since the phase diagrams and various correlation functions for the spin-1 KLM have been analyzed in our previous study\cite{JPSJ.88.024707},
we will use them for comparing with the results for the spin-1/2 KLM.

\subsection{Ground-state phase diagrams}
\label{sub_pd}
In this subsection, we briefly explain the overall features of ground-state phase diagrams for the two KLMs.

\begin{figure}[tb]
	\includegraphics[width=\linewidth]{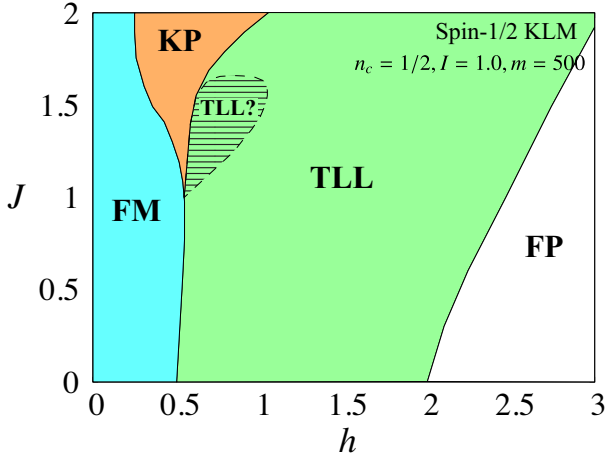}
	\caption{(Color online) Ground-state $h$--$J$ phase diagram of the spin-1/2 KLM for $I=1.0$ and $n_c=1/2$.
	The calculations have been carried out for $N=80$ and $N=120$ with the cutoff $m=500$.
	Each phase is identified by the spin--spin correlation functions $\chi^{\text{c}}_z(r)$ and $\chi^{\ell}_z(r)$ in Eq. (\ref{spin_cf}).
	In the shaded area, $\chi^{\text{c},\ell}_z(r)$ decays exponentially in short range, while it shows a power-law decay in long range.
	The ground state in this area cannot be identified within $N\leq 120$ and $m\leq 500$. 
	}
	\label{S1_PD_I-10}
\end{figure}

\begin{figure}[tb]
	\includegraphics[width=\linewidth]{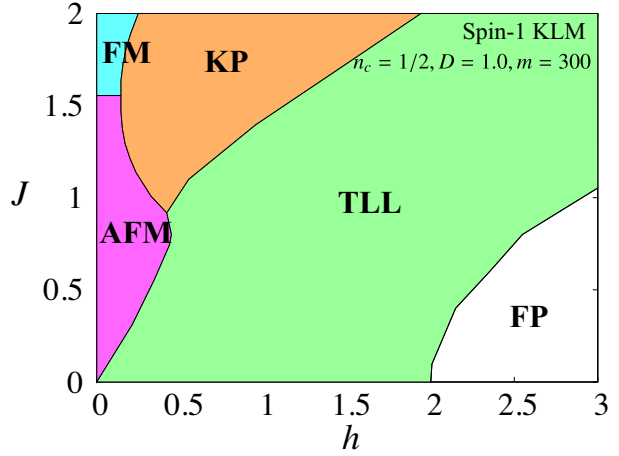}
	\caption{(Color online) Ground-state $h$--$J$ phase diagram of the spin-1 KLM for $D=1.0$ and $n_c=1/2$.
	The calculations have been carried out for $N=80$ and $N=120$ with the cutoff $m=300$.
	Each phase is identified by the spin--spin correlation functions in Eq. (\ref{spin_cf}).
	Other phases diagrams for $n_c\neq 1/2$ are presented in Ref. \citen{JPSJ.88.024707}
	}
	\label{S2_PD_D10}
\end{figure}

Figure \ref{S1_PD_I-10} shows the ground-state phase diagram of the spin-1/2 KLM for 
$I=1.0$ and the electron filling $n_c=1/2$.
The overall features are similar to those of the spin-1 KLM for $D=1.0$ and $n_c=1/2$ as shown in Fig. \ref{S2_PD_D10}.
Each phase is identified by the long-distance behavior of the spin--spin correlation functions:
\begin{align}
	\chi^{\text{c}}_z(r)\coloneqq\Braket{\hat{s}^{z}_{o}\hat{s}^{z}_{o+r}},\;\;
	\chi^{\ell}_z(r)\coloneqq\Braket{\hat{S}^{z}_{o}\hat{S}^{z}_{o+r}}.
	\label{spin_cf}
\end{align}
Here, the bracket represents the ground-state expectation value.
The calculations have been carried out for $(N,o)=(80,20)$ and $(120,30)$.
The basic properties of each phase are as follows.
$\chi^{\text{c},\ell}_z(r)$ remains finite as $r\rightarrow\infty$ in the ferromagnetic (FM) phase, 
decays exponentially in the Kondo plateau (KP) phase, and 
exhibits a power-law decay in the Tomonaga--Luttinger liquid (TLL) phase.
In the fully polarized (FP) phase, the local and electron spins are fully 
polarized by the strong magnetic fields, where $\chi^{\text{c},\ell}_z(r)$ shows an exponential decay.
There is uncertainty in determining the ground state near some phase boundaries.
In the shaded area in Fig. \ref{S1_PD_I-10}, $\chi^{\text{c},\ell}_z(r)$ decays exponentially in short distance,
while it shows a power-law decay in long distance.
We cannot precisely determine the ground state there within our calculations for $N\leq 120$.

When the Kondo coupling $J=0$, the conduction electrons are decoupled from the local spins,
and the FP phase appears at $h_{\text{FP}}=2(1-\cos \pi n_c)$ for the two KLMs.
In addition, for the spin-1/2 KLM, the FM phase appears for $0 \leq h < I/2$,
since the local spin part of the Hamiltonian [Eq. (\ref{S1_HAM})] is the transverse Ising model for $J=0$.

The main difference between the two models is the appearance of AFM phases.
For the spin-1/2 KLM, the FM interaction $I$ between the local spins does not favor AFM correlations and no AFM phases appear.
In contrast, for the spin-1 KLM, the effective interactions for small $J$ are the Ruderman--Kittel--Kasuya--Yosida (RKKY) interactions\cite{RKofRKKY,KofRKKY,YofRKKY},
which lead to the AFM phase with $2k^{0}_{\text F}=\pi n_c$ oscillations \cite{JPSJ.88.024707}.

\subsection{Magnetic properties of the spin-1/2 KLM} \label{sub_mag}
In this subsection, we show numerical results mainly for the spin-1/2 KLM,
focusing on the magnetizations and the spin--spin correlation functions [Eq. (\ref{spin_cf})].

\begin{figure}[bt]
	\centering
	\includegraphics[width=1.0\linewidth]{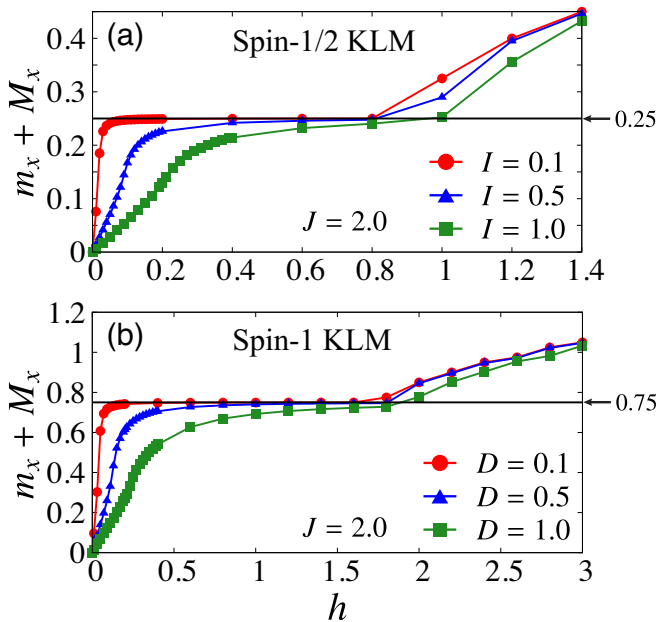}
	\caption{(Color online) Magnetization curves along the $x$-direction for $J=2.0$, $n_c=1/2$ and $N=80$ with the cutoff $m=300$.
	(a) For the spin-1/2 KLM, the ``Kondo plateau" appears at $(1-n_c)/2=0.25$.
	(b) For the spin-1 KLM, the ``Kondo plateau" appears at $1-n_c/2=0.75$.
	The Ising interaction and single-ion anisotropy destroy these plateaus and cause the metamagnetic behavior between the FM and KP phases.
	}
	\label{mag_plat}
\end{figure}

Figure \ref{mag_plat} shows the $h$ dependence of $m_{x}+M_{x}$ for the two KLMs for $N=80$ and $n_c=1/2$.
Here, the magnetizations along the $x$-direction are defined by 
\begin{align}
	m_{x}\coloneqq\frac{1}{N}\sum^{N}_{j=1}\Braket{\hat{s}^{x}_{j}},\;\;
	M_{x}\coloneqq\frac{1}{N}\sum^{N}_{j=1}\Braket{\hat{S}^{x}_{j}}.
	\label{mag_x}
\end{align}
One can see the magnetization plateau (Kondo plateau) at $m^{\text{tot}}_{x}=m_{x}+M_{x}=0.25$ 
for the spin-1/2 KLM [Fig. \ref{mag_plat}(a)]. Similar plateau appears at $m^{\text{tot}}_x=0.75$ 
for the spin 1-KLM as shown in Fig. \ref{mag_plat}(b).

This can be understood as follows.
When $J$ is larger than the other parameters,
all the conduction electrons are tightly coupled to the $S=1/2$ 
local spins and form the spin singlet states. 
The number of remaining local spins is $N(1-n_c)$ and they are 
polarized along the $x$-direction owing to the magnetic field $h$,
which leads to the magnetization plateau at $m^{\text{tot}}_{x}=(1-n_c)/2$.
This is also valid even for the weak-coupling regime as discussed in Refs. \citen{PRL.108.086402} and \citen{PRB.86.165107},
and is related to the notion called spin-selective Kondo insulator.
For the conventional spin-1/2 KLM in the paramagnetic state,
a similar magnetization plateau induced by the longitudinal magnetic fields 
is reported\cite{JPSJ.69.2947,JPSCP.3.011032}. 
Note that the ground state is FM for $h=0$ in our spin-1/2 KLM.
The same mechanism of the magnetization plateau also works for the spin-1 KLM.
All the conduction electrons form the spin-1/2 composites with the $S=1$ local spins,
leading to the magnetization $n_c/2$,
and the remaining local spins, $N(1-n_c)$, are polarized.
In total, the magnetization plateau appears at $m^{\text{tot}}_{x}=(1-n_c)+n_c/2=1-n_c/2$ for large $J$.
Actually, the magnetization plateau at $m=0.75$ can 
be clearly seen for the spin-1 KLM [Fig. \ref{mag_plat}(b)].

\begin{figure}[tb]
	\centering
	\includegraphics[width=\linewidth]{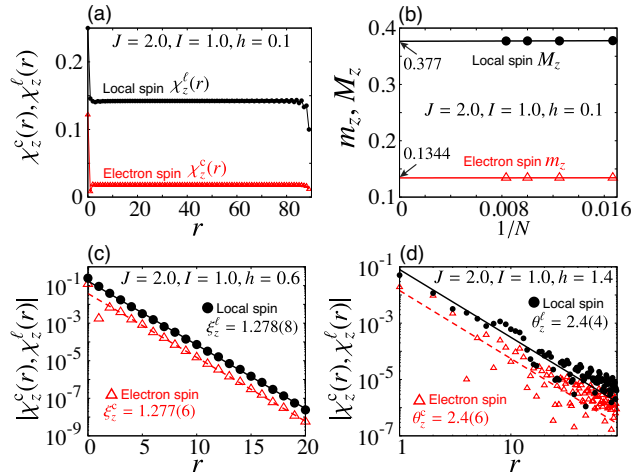}
	\caption{(Color online) 
	Distance $r$ dependences of the spin--spin correlation functions $\chi^{\text{c}}_z(r)$ and $\chi^{\ell}_z(r)$ for (a) $h=0.1$ in the FM phase.
	(b) The system size dependences of magnetizations $m_{z}$ and $M_z$ for $h=0.1$.
	The lines in (b) represent the linear fits for $N=60$, 80, 100, and 120.
	$\chi^{\text{c}}_{z}(r)$ and $\chi^{\ell}_{z}(r)$ for (c) $h=0.6$ in the KP phase and 
	(d) $h=1.4$ in the TLL phase.
	The lines in (c) and (d) represent the linear fits for (c) $11 \leq r \leq 20 $ and (d) $11 \leq r \leq 40$.
	The calculations have been carried out for $J=2.0$, $I=1.0$, $n_c=1/2$ and $N=120$ with the cutoff $m=500$.
	}
	\label{SzSz}
\end{figure}

When $h=I=D=0$, the Hamiltonians in Eqs. (\ref{S1_HAM}) and (\ref{S2_HAM}) have the spin SU(2) symmetries,
and the ground states for the both models are FM for large $J$\cite{PRB.46.13838,PRB.47.2886,PRB.47.8345,PRL.78.2180,PRB.58.2662,JPSJ.88.024707}.
Thus, the magnetization plateaus remain in the zero magnetic field for $I=D=0$.
For finite $I$ or $D$, the Ising anisotropy forces 
the local and electron spins ordered along the $z$-direction,
and the magnetization $m^{\text{tot}}_x$ changes continuously from zero with increasing $h$ as seen in Fig. \ref{mag_plat}.
The reason why $m^{\rm tot}_x$ is larger for smaller $I$ and $D$ in Fig. \ref{mag_plat} is due to the fact that 
the magnetic moment can rotate and is along the $\sim$$x$-direction for smaller $I$ and $D$,
while it is along the $z$-direction for large Ising anisotropy. 
As $I$ and $D$ increase, the magnetization plateaus are gradually smeared out,
and the metamagnetic behavior appears for the small $h$ region as shown in Fig. \ref{mag_plat}.
The phase transition between the FM and KP phases occurs in this region.
For the spin-1 KLM, we have demonstrated in our previous study\cite{JPSJ.88.024707} that this transition belongs to the two-dimensional Ising universality class,
since the interaction between the local spins is Ising-like.
We have confirmed that the same is also true for the spin-1/2 KLM by analyzing critical exponents as expected from the symmetry argument.

Next, we discuss the distance $r$ dependences of the spin--spin correlation functions in Eq. (\ref{spin_cf}) for $n_c=1/2$, $J=2.0$, $I=1.0$, and various $h$ for the spin-1/2 KLM.
Figure \ref{SzSz}(a) shows $\chi^{\text{c},\ell}_z(r)$ as a function of $r$ for $N=120$ and $h=0.1$ with the cutoff $m=500$ in the FM phase.
One can see $\chi^{\text{c},\ell}_z(r) > 0$ for $r\rightarrow\infty$ and this is the signature of the FM state.
We have checked the system size dependences of the magnetizations along the $z$-direction, $m_{z}$ and $M_{z}$, for $N=60$, 80, 100, and 120 as shown in Fig. \ref{SzSz}(b).
Here, we define the order parameters $m_{z}$ and $M_{z}$ in the conventional way:
\begin{align}
	&m_{z}\coloneqq\sqrt{\chi^{\text{c}}_{z}(N/2)},\quad M_{z}\coloneqq\sqrt{\chi^{\ell}_{z}(N/2)}.
	\label{mag_z}
\end{align}
The data in Fig. \ref{SzSz}(b) are obtained by extrapolating $m_{z}$ and $M_{z}$ for $m=400$, 500, 600, and 700 to those for $m=\infty$ for each $N$. The system size and cutoff $m$ dependences are very small, and thus, not shown here. From this analysis, the magnetizations for $h=0.1$ are estimated to be $m_{z}\simeq0.134$ and $M_{z}\simeq0.377$.

In the KP phase, the spin--spin correlation functions decay exponentially, 
and the correlation lengths, $\xi^{\text{c}}_{z}$ and $\xi^{\ell}_{z}$, can be defined by $\chi^{\text{c},\ell}_{z}(r)\sim\exp (-r/\xi^{\text{c},\ell}_{z})$.
This means that there is a finite spin gap corresponding to the finite $\xi^{\text{c},\ell}_{z}$ in the spin sector.
Figure \ref{SzSz}(c) shows the $r$ dependence of $\chi^{\text{c},\ell}_{z}(r)$ for $N=120$, $h=0.6$ and $m=500$ in the KP phase.
The correlation length of the local spin is almost the same as that of the electron spin ($\xi^{\ell}_{z}\simeq\xi^{\text c}_{z}$). This is similar to the results for the spin-1 KLM.\cite{JPSJ.88.024707}
We have checked the system size dependence up to $N=120$ and the cutoff $m$ dependence up to $m=700$,
and they turn out to be sufficiently small.

For the TLL phase, the spin--spin correlation functions 
show a power-law decay $\chi^{\text{c},\ell}_{z}(r)\sim r^{-\theta^{\text{c},\ell}_{z}}$,
which means that the system is critical.
Figure \ref{SzSz}(d) represents that $r$ dependence of $\chi^{\text{c},\ell}_{z}(r)$ in the TLL phase shows a power-law decay $\chi^{\text{c},\ell}_z(r)\sim r^{-\theta^{\text{c},\ell}_z}$.
Unfortunately, $\chi^{\text{c},\ell}_{z}(r)$ seems to show oscillations. Owing to this, 
the precise estimation of the critical exponents, $\theta^{\text{c}}_{z}$ and 
$\theta^{\ell}_{z}$, is difficult within our calculations for the system size 
up to $N=120$ and the cutoff $m=700$.

\section{Superconducting Correlations}
\label{sec_sc}
In this section, we will discuss the SC correlations in detail.
The conventional SC order parameter is defined by the Cooper pair amplitude:
\begin{equation}
	\Phi_{\sigma\sigma'}(i,j)=\Braket{\hat{c}_{i,\sigma}\hat{c}_{j,\sigma'}}.\label{eq:Phi0}
\end{equation}
This is the order parameter breaking the charge U(1) symmetry,
and $\Phi_{\sigma\sigma'}(i,j)$ is classified into the spin-singlet or the spin-triplet in spin-SU(2) symmetric systems. 
The fermion antisymmetry requires that the spin-singlet (spin-triplet) Cooper pairs be even-parity (odd-parity) when the two spatial coordinates are interchanged $i\leftrightarrow j$. 
The more sophisticate analyses taking into account the symmetry of the lattice,\cite{RMP.63.239,PRB.97.134512} orbitals,\cite{PRB.94.174513} and topological aspects,\cite{PRB.97.134512,PRB.90.165114,arXiv:1909.09634} have been developed.
There is another way to extend such classification and it was studied by Berezinskii\cite{JETPL.20.287}, where the properties under the time (or frequency) domain are taken into account.
There, odd-frequency superconductivity\cite{arXiv1709.03986} was proposed.
One way to generate the odd-frequency superconductivity is known to construct a ``composite pair'' consisting of electrons and some bosons as\cite{PRB.48.7445,PRB.52.1271,Dahal_2009}
\begin{equation}
	\Phi^{\mathcal O_k}_{\sigma\sigma'}(i,j)=\Braket{\hat{c}_{i,\sigma}\hat{c}_{j,\sigma'}{\mathcal O}_k}. \label{eq:Phi1}
\end{equation}
\tred{In the one-dimensional KLM with anisotropy in the Kondo couplings, the bosonization analysis demonstrated that such composite pair correlations become critical under the spin gapped phase.\cite{PhysRevLett.77.1342}}
Hoshino and Kuramoto in their DMFT study\cite{PRL.112.167204} have demonstrated that the two-channel KLM indeed shows the odd-frequency superconductivity of this type with $\mathcal O\sim \hat{\bm S}$.
Recently, similar composite superconductivity has been discussed for the case of semimetallic conduction bands\cite{PRB.100.094532}.
In this respect, for the KLMs,
it is important to take into account such composite Cooper pairs for discussing the superconductivity,
\tred{since there are finite couplings between the conventional Cooper pair [Eq. (\ref{eq:Phi0})] and the composite one [Eq. (\ref{eq:Phi1})] if $\mathcal O_k$ is appropriately chosen.
In one-dimensional systems, the correlation function corresponding to the conventional Cooper pair [Eq. (\ref{eq:Phi0})] has been examined to clarify the SC correlations by using the DMRG method in various systems.\cite{JPSJ.75.043710,PhysRevB.83.205113,PhysRevB.79.220513,PhysRevB.98.140505,Jiang1424}
In this study, we will analyze generalized Cooper pairs including both the conventional and the composite ones by the DMRG.}
It should be noted that our target is not the odd-frequency superconductivity but the composite SC order parameters. 
Since the data in the (static) DMRG are those for the equal-time quantity, 
if the odd-frequency superconductivity is realized, it is 
 reflected in the corresponding equal-time composite order parameters.

In the following, we will discuss the SC correlations including such composite pairs mainly for the spin-1/2 KLM. We will first show the formulation of the generalized Cooper pairs for our models in Sect. \ref{sec_form}.
Then, we will show the numerical results of the SC correlations in Sect. \ref{sec_pow}.
The detailed profile of the composite Cooper pairs \tblue{is} discussed in Sect. \ref{sec_coop}.
We will briefly discuss the results for the spin-1 KLM in Sect. \ref{secKLM1}.


\subsection{Composite Cooper pairs}
\label{sec_form}
In the KLMs, the local spins and electrons strongly interact with each other,
and thus, there is no reason to ignore the composite Cooper pairs when one discusses the superconductivity.
We consider the composite Cooper pairs between the local spins and the electrons in addition to the conventional pairs within the nearest neighbors.
Although the restriction of the size of the pairs is simply due to the computational costs,
we believe that it is sufficient when we discuss the relative amplitude between the composite and non-composite pairs.
Within this restriction, we take into account the complete operator bases including
the first-order composite $\Phi^{\mathcal O_{j}}_{\sigma\sigma'}(j,j)=\braket{\hat{c}_{j,\sigma}\hat{c}_{j,\sigma'}{\mathcal O}_{j}}$
and the second-order one $\Phi^{{\mathcal O}_i{\mathcal O}_j}_{\sigma\sigma'}(i,j)=\braket{\hat{c}_{i,\sigma}{\mathcal O}_i\hat{c}_{j,\sigma'}{\mathcal O}_j}$ for $i\ne j$, with 
\begin{equation}
	{\mathcal O}_j=\ket{\alpha}_j\bra{\beta}_j.
\end{equation}
\tred{ Since there is a linear combination of $\mathcal O_j$'s that is equal to the identity operator $\sum_\alpha |\alpha\rangle_j \langle \alpha|_j$, these composite pair amplitudes include the conventional Cooper pair amplitude. }
Note that we retain only operators that the electron and local spin couple locally, i.e., we have ignored 
ones such as $\Braket{\hat{c}_{i,\sigma}\hat{c}_{i,\sigma'}{\mathcal O}_{i+1}}$. 

In actual calculations, we use the spin-inversion parity basis as summarized in Appendix \ref{append}. We represent the generalized Cooper-pair operators in the spin-inversion parity basis as (rewrite $\Phi$ as $\Delta$)
\begin{align}
	\hat{\Delta}^{\alpha\beta}_{pq}(j,k)&\coloneqq
	\hat{c}_{j,p}{\mathcal O}_j^\alpha  \hat{c}_{k,q}{\mathcal O}_k^\beta,\quad (j\ne k), 
	\label{Cooper_in}\\
	\hat{\Delta}^{\alpha}_{-+}(j,j)&\coloneqq
	\hat{c}_{j,-}\hat{c}_{j,+}{\mathcal O}_j^\alpha,
	\label{Cooper_on}
\end{align}
where, $\alpha$ and $\beta$ run from 1 to 4 (9) for the spin-1/2 (spin-1) KLM and ${\mathcal O}_j^\alpha$'s are listed in Table \ref{tbl-1}. $p,q=\pm$ represent the spin-inversion parity. 
For the spin-1/2 KLM, the local spin bases with the eigenvalue of the spin-inversion parity $p=\pm$ are defined as
\begin{equation}
	\ket{\pm}_j\coloneqq \frac{1}{\sqrt{2}}\left(\ket{\uparrow}_j \pm \ket{\downarrow}_j \right).\label{eq:Defpm}
\end{equation}
For the spin-1 KLM, see Appendix\ref{append}. Similarly, the electron annihilation operator in the spin-inversion parity bases are also defined as 
\begin{equation}
	\hat{c}_{j,\pm}\coloneqq \frac{1}{\sqrt{2}}(\hat{c}_{j, \uparrow}\pm \hat{c}_{j, \downarrow} ).
	\label{eq:Defpm2}
\end{equation}

\begin{table}[t!]
	\caption{Definitions of the local operators $\mathcal{O}^\alpha_j$ for the spin-1/2 KLM and spin-1 KLM.
	$p$ represents the eigenvalue of the spin-inversion parity.
	The site index $j$ is suppressed for simplicity in the list. For the definition of the local spin states for the spin-1/2 KLM, see Eq. (\ref{eq:Defpm}).
	The local spin states for the spin-1 KLM are defined in Appendix \ref{append}.}\vspace{3mm}
	\begin{center}
		\begin{tabular}{llllllllll}
			\hline\hline
			\multicolumn{3}{c}{spin-1/2 KLM} & \hspace{1cm}&
			\multicolumn{6}{c}{spin-1 KLM}\\
			$\alpha$& \multicolumn{1}{c}{${\mathcal O}^\alpha_j$} &$p$&&$\alpha$ & \multicolumn{1}{c}{${\mathcal O}^\alpha_j$} &$p$&$\alpha$ & \multicolumn{1}{c}{${\mathcal O}^\alpha_j$} &$p$\\
			\hline
			$1$& $\ket{+} \bra{+}$ &$+$&&$1$& $\ket{+}\bra{+}$ & $+$&$5$& $\ket{0} \bra{0}$ & $+$\\
			$2$& $\ket{+} \bra{-}$ &$-$&&$2$& $\ket{+}\bra{0}$ & $+$&$6$& $\ket{0} \bra{-}$ & $-$\\
			$3$& $\ket{-} \bra{+}$ &$+$&&$3$& $\ket{+}\bra{-}$ & $-$&$7$& $\ket{-} \bra{+}$ & $-$\\
			$4$& $\ket{+} \bra{+}$ &$-$&&$4$& $\ket{0}\bra{+}$ & $+$&$8$& $\ket{-} \bra{0}$ & $-$\\
			& &&&& & &$9$& $\ket{-} \bra{-}$ & $+$\\
			\hline\hline
		\end{tabular}
	\end{center}
	\label{tbl-1}
\end{table}
\noindent Since one can take ${\mathcal O}_j^\alpha$ as an operator with a definite 
spin-inversion parity similarly to $\hat{c}_{j,p}$ as shown in Table \ref{tbl-1},  
$\hat{\Delta}^{\alpha\beta}_{p,q}(j,k)$ and $\hat{\Delta}^{\alpha}_{-,+}(j,j)$ also 
have definite parity eigenvalues. The lack of the SU(2) spin symmetry in our models inevitably leads to the classification based on the spin-inversion parity.
Note also that $\hat{\Delta}^{\alpha}_{-,+}(j,j)=-\hat{\Delta}^{\alpha}_{+,-}(j,j)$
and $\hat{\Delta}^{\alpha}_{+,+}(j,j)=\hat{\Delta}^{\alpha}_{-,-}(j,j)=0$. Thus, it is sufficient to consider
 only $\hat{\Delta}^{\alpha}_{-,+}(j,j)$ for the on-site pairs.

In terms of the generalized Cooper pairs in Eqs. (\ref{Cooper_in}) and (\ref{Cooper_on}),
we define the SC correlation functions as 
\begin{align}
	\chi^{\alpha;\beta}(i,i;j,j)
	&\coloneqq\Braket{\left[\hat{\Delta}^\alpha_{-+}(i,i)\right]^{\dagger}\hat{\Delta}^\beta_{-+}(j,j)},\label{sccf_oo}\\
	\chi^{\alpha;\beta\gamma}_{-+;pp'}(i,i;j,k)
	&\coloneqq\Braket{\left[\hat{\Delta}^{\alpha}_{-+}(i,i)\right]^{\dagger}\hat{\Delta}^{\beta\gamma}_{pp'}(j,k)},\label{sccf_oi}\\
	\chi^{\alpha\beta;\gamma}_{pp';-+}(i,j;k,k)
	&\coloneqq\Braket{\left[\hat{\Delta}^{\alpha\beta}_{pp'}(i,j)\right]^{\dagger}\hat{\Delta}^\gamma_{-+}(k,k)},\label{sccf_io}\\
	\chi^{\alpha\beta;\gamma\delta}_{pp';qq'}(i,j;k,l)
	&\coloneqq\Braket{\left[\hat{\Delta}^{\alpha\beta}_{pp'}(i,j)\right]^{\dagger}\hat{\Delta}^{\gamma\delta}_{qq'}(k,l)}.\label{sccf_ii}
\end{align}
For some Cooper pairs, these SC correlation functions are equal to zero. This is due to the conservation of the spin-inversion parity eigenvalue. Thus, there are two decoupled (block-diagonalized) sectors. One is the even parity $\hat{\Delta}$ and the other is the odd parity $\hat{\Delta}$. Any correlation functions which consist of odd numbers of the parity-odd operators vanish. Thus, we just need to calculate the correlation functions in the two sectors separately.

\begin{figure*}[h]
	\centering
	\includegraphics[width=\linewidth]{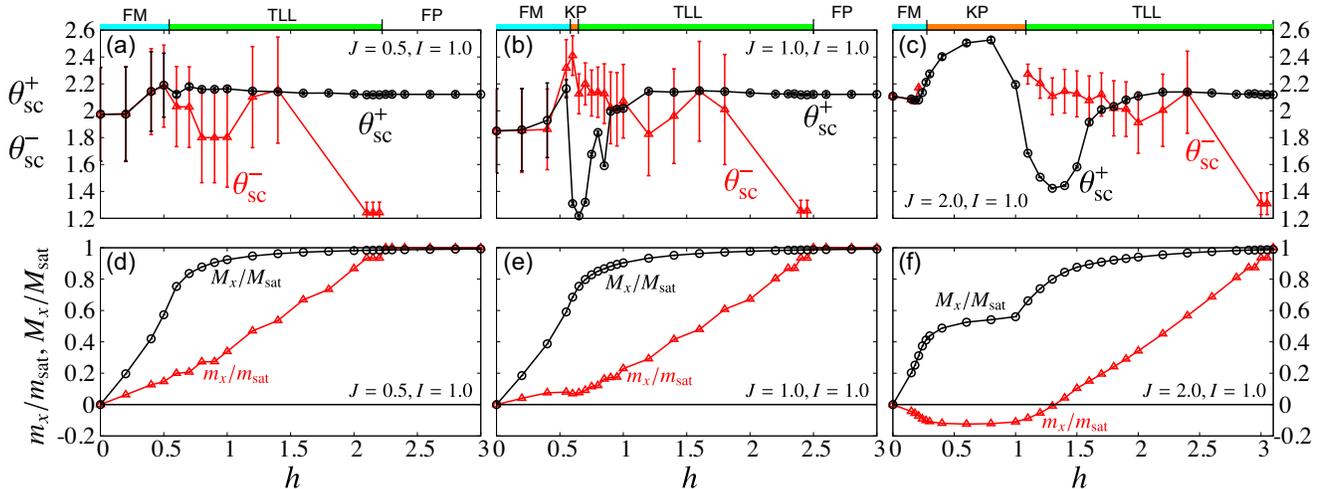}	
	\caption{(Color online)
	Magnetic field $h$ dependences of the exponent $\theta^{\pm}_{\text{sc}}$ and the magnetizations $m_x$ and $M_x$ along the $x$-direction for the spin-1/2 KLM.
	The calculations have been carried out for $I=1.0$, $n_c=1/2$, and $N=120$ with the cutoff $m=500$.
	(a) $\theta^{\pm}_{\text{sc}}$ for $J=0.5$,
	(b) $\theta^{\pm}_{\text{sc}}$ for $J=1.0$,
	and (c) $\theta^{\pm}_{\text{sc}}$ for $J=2.0$.
	Note that $\theta^{-}_{\text{sc}}\simeq 1.2$ in the TLL phase close to the FP phase does not mean the enhancement of the SC correlations.
	See the main text.
	The magnetizations $m_x$ and $M_x$ are shown for (d) $J=0.5$, (e) $J=1.0$, and (f) $J=2.0$, with $m_{\text{sat}}=0.25$ and $M_{\text{sat}}=0.5$.
	The enhancement of the SC correlations characterized by $\theta^{+}_{\text{sc}}<2$ occurs when $m_x\simeq 0$.
		}
	\label{S1Sc}
\end{figure*}

To study the composite-pairing superconductivity within nearest neighbors,
we introduce the two SC correlation matrices for the even ($p=+$) and odd ($p=-$) sectors as
\begin{align}
	\hat{\chi}^{p}_{\text{sc}}(k,l)\coloneqq
	\begin{pmatrix}
	\hat{\chi}^{p}_{k,k;l,l}     & \hat{\chi}^{p}_{k,k;l,l+1}     & \hat{\chi}^{p}_{k,k;l+1,l+1}  \\
	\hat{\chi}^{p}_{k,k-1;l,l}   & \hat{\chi}^{p}_{k,k-1;l,l+1}   & \hat{\chi}^{p}_{k,k-1;l+1,l+1}\\
	\hat{\chi}^{p}_{k-1,k-1;l,l} & \hat{\chi}^{p}_{k-1,k-1;l,l+1} & \hat{\chi}^{p}_{k-1,k-1;l+1,l+1}
	\end{pmatrix}.
	\label{sc_cf}
\end{align}
Here, for example, the matrix elements of $\hat{\chi}^p_{k,k;l,l}$, $\hat{\chi}^p_{k,k;l,l+1}$, and $\hat{\chi}^p_{k,k-1;l,l+1}$ can be read from $\chi^{\alpha;\beta}(k,k;l,l)$ in Eq. (\ref{sccf_oo}),  $\chi^{\alpha;\beta\gamma}_{-+;pp'}(k,k;l,l+1)$ in Eq. (\ref{sccf_oi}), and $\chi^{\alpha\beta;\gamma\delta}_{pp';qq'}(k,k-1;l,l+1)$ in Eq. (\ref{sccf_ii}) with the spin-inversion parity $p$, respectively.

We diagonalize the SC correlation matrices $\hat{\chi}^{p}_{\text{sc}}(k,k+r)$ in Eq. (\ref{sc_cf}),
and obtain the eigenvalues $\lambda^{p}_{\text{sc}}(r)$ and the corresponding eigenvectors $\bm{v}^{p}_{\text{sc}}(r)$. Here, we set $k=N/2+1$ (the middle of the system) to reduce the boundary effects.
The strongest SC correlation corresponds to the maximum eigenvalue of $|\lambda^{\pm}_{\text{sc}}(r)|$ and the corresponding eigenvector.
We denote them as $\lambda^{p}_{\text{max}}(r)$ and $\bm{v}^{p}_{\text{max}}(r)$.
The SC correlations are characterized by a power-law decay:
\begin{align}
	|\lambda^{p}_{\text{max}}(r)|\sim r^{-\theta^{p}_{\text{sc}}}, \quad (p=\pm), 	\label{sc_pow}
\end{align}
or by an exponential decay:
\begin{align}
	|\lambda^{p}_{\text{max}}(r)|\sim \exp(-r/\xi^{p}_{\text{sc}}),  \quad (p=\pm).
\end{align}
By analyzing the exponents $\theta^{\pm}_{\text{sc}}$ or the correlation length $\xi^{\pm}_{\text{sc}}$,
we can discuss the strength of the SC correlations.
Note that since $\theta^{\text{free}}_{\text{sc}}=2$ for the one-dimensional free electrons,
$\theta^{p}_{\text{sc}} < 2$ is the signature of the enhanced SC correlations,
and the corresponding eigenvectors characterize the nature of the Cooper pairs.

\tred{We emphasize again that the eigenvectors do not necessarily consist of purely conventional Cooper pairs or ``genuine'' composite pairs (i.e. the linear combination of $\mathcal O$'s does not include the identity),
but the mixture of them allowed by the symmetry of the system.
This is simply because there are finite matrix elements in Eq. (\ref{sc_cf}) among them.
Clarifying such profile of the Cooper pairs is one of our main interests in this paper.
For local Cooper pairs as in the DMFT, the conventional one is a spin-SU(2) singlet 
$\langle \hat{c}_{j\uparrow}\hat{c}_{j\downarrow}\rangle$, while the composite one is a 
spin-SU(2) triplet  
$\langle \hat{c}_{j\uparrow}\hat{c}_{j\downarrow}\hat{S}_j^{x,y,z}\rangle$. Thus, they never hybridize. If $\hat{S}_j$ is replaced by some spin-SU(2) singlet operators, then the two can hybridize. Note that the situation for our models is similar to the latter case.
}

Before discussing the numerical results,
let us comment on a subtle issue about the SC correlation matrices in Eq. (\ref{sc_cf}).
In the open boundary condition (OBC), the SC correlation matrices are not symmetric,
which means that the eigenvalues can be complex values
and the right eigenvectors $\bm{v}_{\text{R}}(r)$ are not equal to the left eigenvectors $\bm{v}_{\text{L}}(r)$.
Considering the fact that the SC correlation matrices are symmetric, if the translational symmetry is present, 
 the antisymmetric parts of the SC correlation matrices are expected to be negligibly small even in the OBC,
as long as the system size is sufficiently large.
In our calculations, we take the system size up to $N=120$,
and we find that $\lambda^{p}_{\text{max}}(r)$ takes a real value for any $r$.
For the eigenvectors, the elements can always be taken real,
and $\bm{v}^{+}_{\text{R},\text{max}}(r)\simeq [\bm{v}^{+}_{\text{L},\text{max}}(r)]^{\mathsf T}$ for all the phases.
There are some cases where $\bm{v}^{-}_{\text{R},\text{max}}(r)$ deviates from $[\bm{v}^{-}_{\text{L},\text{max}}(r)]^{\mathsf T}$.
However, in these cases, $|\lambda^{+}_{\text{max}}(r)|$ is larger than $|\lambda^{-}_{\text{max}}(r)|$
and the Cooper pairs corresponding to $\bm{v}^{-}_{\text{R},\text{max}}(r)$ is not the leading eigenmode.
Thus, this does not affect our main conclusions.

\begin{figure*}[t!]
	\centering
	\includegraphics[width=1.0\linewidth]{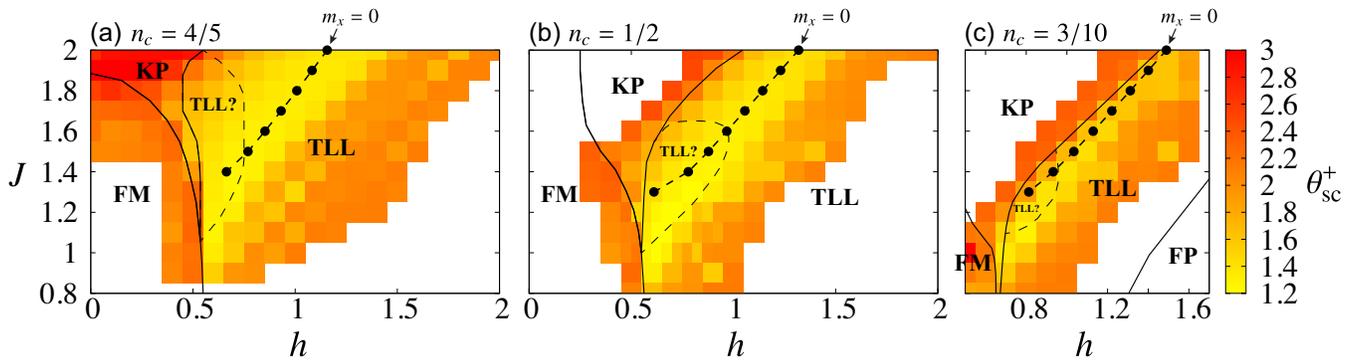}	
	\caption{(Color online)
	$\theta^{+}_{\text{sc}}$ as functions of $h$ and $J$ for the spin-1/2 KLM.
	The calculations have been carried out for $N=120$ and $I=1.0$ with the cutoff $m=500$.
	(a) $n_c=4/5$, (b) $n_c=1/2$, and (c) $n_c=3/10$.
	$\theta^{+}_{\text{sc}}$ is obtained by the power-law fitting of $\lambda^{+}_{\text{max}}(r)$ from  $r=11$ to $30$.
	The dashed lines with filled circles represent the curves along which $m_x\simeq 0$.
	It is difficult to determine the ground state in the areas surrounded by the dashed line.
	See Fig. \ref{S1_PD_I-10}.
	}
	\label{sc_phase}
\end{figure*}

\subsection{Numerical results: superconducting correlations}
\label{sec_pow}
In this subsection, we show the numerical results of the SC correlations,
 focusing on $\theta^{\pm}_{\text{sc}}$.
Figure \ref{S1Sc} shows the $h$ dependences of the SC exponents $\theta^{\pm}_{\text{sc}}$ [Eq. (\ref{sc_pow})],
and the magnetizations along the $x$-direction $m_{x}/m_{\text{sat}}$ and $M_{x}/M_{\text{sat}}$ [Eq. (\ref{mag_x})].
Here, the saturation moments are $m_{\text{sat}}=n_{c}/2=0.25$ and $M_{\text{sat}}=0.5$.
The calculations have been carried out for the system size $N=120$,
the electron filling $n_c=1/2$, the Ising interaction $I=1.0$,
and the Kondo coupling $J=0.5,\:1.0,\:2.0$ with the cutoff $m=500$.
It is noted that the SC correlations corresponding to $\lambda^{+}_{\text{max}}(r)$ are strongly enhanced, i.e., $\theta^+_{\rm sc}<2$ in the TLL phase close to the KP phase,
while the SC correlations corresponding to $\lambda^{-}_{\text{max}}(r)$ are not enhanced in all the phases with relatively larger errors as shown in Figs. \ref{S1Sc}(a)-\ref{S1Sc}(c).

Let us start by discussing $\theta^{\pm}_{\text{sc}}$ in the FM phase.
One can see $\theta^{+}_{\text{sc}}\simeq\theta^{-}_{\text{sc}}\sim 2$ within the error bars [Figs. \ref{S1Sc}(a)-\ref{S1Sc}(c)].
The fitting errors are slightly larger for $J=0.5$ and $J=1.0$ [Figs. \ref{S1Sc}(a) and \ref{S1Sc}(b)] than those for $J=2.0$.
Comparing the data with the free-electron case: $\theta^{\text{free}}_{\text{sc}}=2$, one can say that the SC correlations are not so enhanced in the FM phase.
For much larger $J$ limit, the SC correlations become enhanced.
This will be discussed in Sect. \ref{sec_dis_compare}.

For the KP phase, there is the energy gap between the ground state with the total spin-inversion parity $P=1$ and the excited states with $P=-1$.
This leads to the exponential decay of $\lambda^{-}_{\text{sc}}(r)\sim \exp(-r/\xi^{-}_{\text{sc}})$,
since $\hat{\chi}^{-}_{\text{sc}}(k,k+r)$ consists of the odd spin-inversion parity Cooper pairs.
Thus, $\theta^{-}_{\text{sc}}$ cannot be defined, 
and the corresponding data for $\theta^{-}_{\text{sc}}$ in Fig. \ref{S1Sc}(c) are absent.
$\lambda^{+}_{\text{sc}}(r)$ shows a power-law decay 
and $\theta^{+}_{\text{sc}}>2$.
These observations indicate that 
the SC correlations are suppressed in the KP phase. This is consistent with the physical picture of the KP phase, where the electrons and the local spins are strongly bound with a finite excitation gap.

In clear contrast to the above, the SC correlations corresponding to $\lambda^{+}_{\text{sc}}(r)$ are strongly enhanced in the TLL phase close to the KP phase as shown in Fig. \ref{S1Sc}.
In particular, $\theta^{+}_{\text{sc}}\sim 1.2$ for $J=1.0$ and $h=0.65$ [Fig. \ref{S1Sc}(b)],
where the KP phase fades out and the FM phase is also suppressed by $h$.
$\theta^{+}_{\text{sc}}\sim 1.2$ is the minimum value \tblue{throughout} our calculations for $I=1.0$ and $0\leq J \leq 2.0$.
Away from the KP phase with increasing $h$,
$\theta^{+}_{\text{sc}}$ returns to the larger value $\gtrsim 2$. 
This tendency is also seen for $J=2$ as shown in Fig. \ref{S1Sc}(c), and $\theta^+_{\rm sc}\sim 1.4$ for $h\simeq 1.4$ close to the KP phase.
For smaller $J=0.5$, such enhanced SC correlations are absent as shown in Fig. \ref{S1Sc}(a),
where there is no KP phase.
The relation between the enhanced SC and the KP phase will be discussed later in this subsection.

For the SC correlations corresponding to $\lambda^{-}_{\text{sc}}(r)$, 
owing to the large errors in $\lambda^{-}_{\text{sc}}(r)$, 
we cannot estimate $\theta^{-}_{\text{sc}}$ for the TLL phase in the high-field region,
since the power-law fitting does not work. 
The difficulty in the fitting can be understood as follows.
Let us consider a simple case: one-dimensional free fermions with the transverse magnetic field $h$.
In this case, the SC correlation function for the odd parity sector with the periodic boundary condition is given by
\begin{align}
	\Braket{\hat{c}^{\dagger}_{j,+}\hat{c}^{\dagger}_{j,-}\hat{c}_{k,-}\hat{c}_{k,+}}=\frac{\sin [k^{\text{up}}_{\text{F}}(k-j)]\sin [k^{\text{low}}_{\text{F}}(k-j)]}{\pi^2(k-j)^2}.
\end{align}
Here, $k^{\text{up}}_{\text{F}}$ and $k^{\text{low}}_{\text{F}}$ are the Fermi wavenumbers of the upper and lower conduction bands, respectively, which are split by $h$.
As $h$ approaches $h_{\text{FP}}=2(1-\cos\pi n_c)$, $k^{\text{up}}_{\text{F}}$ and $k^{\text{low}}_{\text{F}}$ approach to $0$ and $\pi n_c$, respectively with the condition $k^{\text{up}}_{\text{F}}+k^{\text{low}}_{\text{F}}=\pi n_c$ kept.
The presence of the long range oscillation $\lambda^{\text{up}}_{\text{F}}=2\pi/k^{\text{up}}_{\text{F}}$ appearing in the high field-region is the reason why the power-law fitting does not work for our finite size calculations.
In addition to this, the situation in actual calculations is more complicated, since there exist the boundary effects from the open edges.

Let us return to the data $\theta^{-}_{\text{sc}}$.
One can notice $\theta^{-}_{\text{sc}}\sim 1.2$ in the TLL phase close to the FP phase.
However, this does not mean the enhancement of the SC correlations.
In this region, the system size is too small to estimate the SC exponent $\theta^{-}_{\text{sc}}$ correctly and
the value $\theta^{-}_{\text{sc}}\sim 1.2$ is not reliable.
This is also understood by the free fermion case.
If $k^{\text{up}}_{\text{F}}$ is very small and the length $|k-j|$ is not sufficiently large,
\begin{align}
	\frac{\sin [k^{\text{up}}_{\text{F}}(k-j)]\sin [k^{\text{low}}_{\text{F}}(k-j)]}{\pi^2(k-j)^2}\sim\frac{k^{\text{up}}_{\text{F}}\sin [k^{\text{low}}_{\text{F}}(k-j)]}{\pi^2(k-j)}.
	\label{kf}
\end{align}
This leads to underestimation of $\theta^{-}_{\text{sc}}$.

In the FP phase, the SC correlations are trivial.
The conduction electrons are decoupled from the local spins owing to the strong magnetic field,
which leads to $\theta^{+}_{\text{sc}} \simeq \theta^{\text{free}}_{\text{sc}}=2$.
As in the KP phase, there is a finite energy gap between the ground state and the excited state with the distinct spin-inversion parity from that of the ground state.
Thus, $\lambda^{-}_{\text{sc}}(r) \sim \exp(-r/\xi^{-}_{\text{sc}})$ and $\theta^{-}_{\text{sc}}$ is not shown in Figs. \ref{S1Sc}(a) and \ref{S1Sc}(b).

To summarize the above results, the SC correlations corresponding to $\lambda^{+}_{\text{max}}(r)$ are enhanced ($\theta^{+}_{\text{sc}} < 2$) in the TLL phase close to the KP phase,
suppressed ($\theta^{+}_{\text{sc}} > 2$) in the KP phase, and neither enhanced nor suppressed in other phases.
These are our main results in this paper.
For the SC correlations corresponding to $\lambda^{-}_{\text{max}}(r)$,
they show the exponential decay in the KP phase,
and $\theta^{-}_{\text{sc}} \gtrsim 2.0$ for the other phases.
There is no clear sign of the enhancement of the SC correlations corresponding to $\lambda^{-}_{\text{max}}(r)$.

Now, we discuss the relation between the enhanced SC correlations and the magnetization $m_{x}$.
For large $J$ and small $h$, the system is in the FM phase.
There, the alignment of the electron spins is antiparallel to the local spins ordered along the $z$-direction owing to $I>0$.
Thus, the electron spins are aligned along the $z$-direction.
If $h$ is small, both the electron and local spins slightly tilt to the $x$-direction,
leading to small negative value $m_{x}<0$ in the KP phase [Fig. \ref{S1Sc}(f)].
As $h$ increases, $m_{x}$ also increases and is equal to zero at an intermediate field.
The enhancement of the SC correlations corresponding to $\lambda^{+}_{\text{max}}(r)$ occurs around this region,
which means that the vanishing effective magnetic fields for the conduction electrons is strongly related to the enhancement of the SC correlations.
The similar tendency is seen for the $J=1.0$ case [Fig. \ref{S1Sc}(e)],
while the SC correlations are not enhanced for $J=0.5$, since $m_{x}$ shows a monotonic increase as shown in Fig. \ref{S1Sc}(d) and this is related to the absence of the KP phase.

Let us finally discuss the relation between the SC correlations and the phase diagram for the spin-1/2 KLM.
Figures \ref{sc_phase}(a)-\ref{sc_phase}(c) represent $\theta^{+}_{\text{sc}}$ for $n_c=4/5$, $1/2$, and $3/10$, respectively.
One can see that the SC correlations are enhanced in the TLL phase close to the KP phase.
The condition for the enhancement is indeed related to the vanishing conduction electron magnetization $m_x\sim 0$.
In Fig. \ref{sc_phase}, in addition to the phase boundary among the KP, FM, TLL, and FP phases,
the curve along which $m_x\sim 0$ is also indicated by dashed lines with filled circles.
Clearly, $\theta^{+}_{\rm sc}$ shows a valley structure along which $m_x\sim 0$ for all the three filling cases.
Thus, the tendency is common for the wide range of parameter spaces.
Note that, for smaller fillings, the SC correlations are gradually suppressed as seen in Fig. \ref{sc_phase}(c).
\subsection{Profile of composite Cooper pairs}\label{sec_coop}

\begin{figure*}[t]
	\centering
	\includegraphics[width=\linewidth]{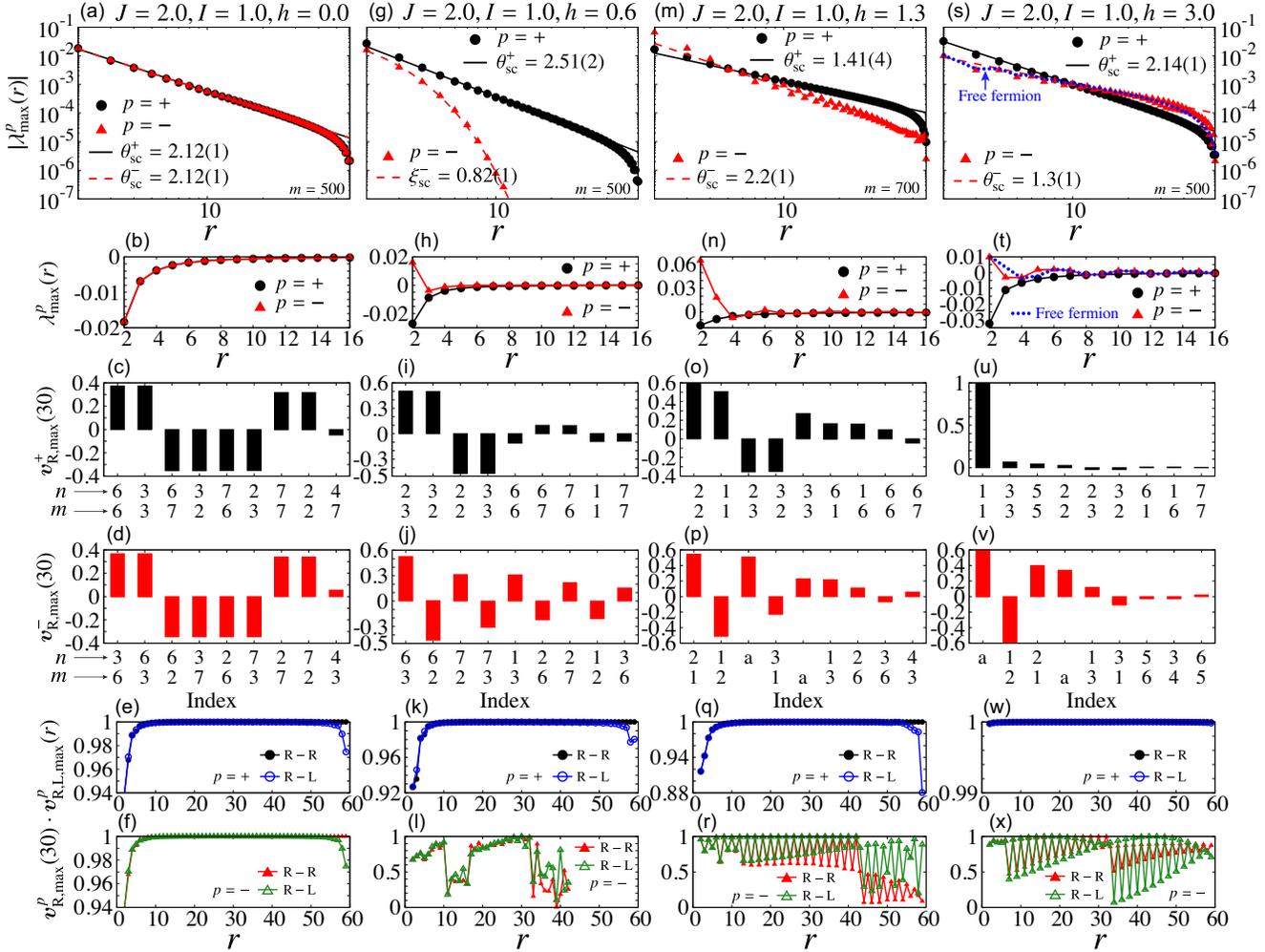}	
	\caption{(Color online)
	Distance $r$ dependences of the maximum eigenvalue $\lambda^{\pm}_{\text{max}}(r)$ and corresponding eigenvectors $\bm{v}^{\pm}_{\text{R},\text{max}}(30)$ for $I=1.0$, $J=2.0$, $n_c=1/2$, and $N=120$.
	The results for (a)-(f) the FM phase with $h=0.0$ and the cutoff $m=500$,
	(g)-(l) the KP phase with $h=0.6$ and $m=500$,
	(m)-(r) the TLL phase with $h=1.3$ and $m=700$,
	and (s)-(x) the TLL phase with $h=3.0$ and $m=500$. 
	(a), (g), (m), and (s) $r$ dependences of $|\lambda^{\pm}_{\text{max}}(r)|$,
	(b), (h), (n), and (t) short-range behavior of $\lambda^{\pm}_{\text{max}}(r)$,
	(c), (i), (o), and (u) the largest nine elements in $\bm{v}^{+}_{\text{R},\text{max}}(30)$,
	(d), (j), (p), and (v) the largest nine elements in $\bm{v}^{-}_{\text{R},\text{max}}(30)$.
	The two numbers and an alphabet for the horizontal axis show the indices $n$ and $m$ in Eq. (\ref{eq:S1}) or $n$ in Eq. (\ref{eq:S2}).
	One can read the corresponding Cooper pairs from Table \ref{list_coop}.
	(e), (k), (q), and (w) $r$ dependences of the inner products $\bm{v}^{+}_{\text{R},\text{max}}(30)\cdot\bm{v}^{+}_{\text{R},\text{max}}(r)$
	and $\bm{v}^{+}_{\text{R},\text{max}}(30)\cdot[\bm{v}^{+}_{\text{L},\text{max}}(r)]^{\mathsf{T}}$,
	and (f), (l), (r), and (x) $r$ dependences of the inner products $\bm{v}^{-}_{\text{R},\text{max}}(30)\cdot\bm{v}^{-}_{\text{R},\text{max}}(r)$ and $\bm{v}^{-}_{\text{R},\text{max}}(30)\cdot[\bm{v}^{-}_{\text{L},\text{max}}(r)]^\mathsf{T}$. 
	In (s), the result for the free fermions is indicated by the dotted line.
	See Eq. (\ref{kf}).
	}
	\label{sc_comp}
\end{figure*}

In this subsection, we will discuss the profile of the Cooper pairs. The eigenvector of 
Eq. (\ref{sc_cf}), 
$\bm{v}_{\text{R},\text{max}}^{\pm}(r)$, contains the information about which components are dominant.
The corresponding eigenvalue $\lambda^{\pm}_{\text{max}}(r)$ represents the amplitude of the correlations at the distance $r$. Since the diagonalization of the matrix [Eq. (\ref{sc_cf})] is carried out for each $r$ separately, the eigenvectors can vary as functions of $r$. As a natural expectation, one expects the long-distance correlations are dominated by one mode (or several if spatial oscillations are present). 
The inner products, $\bm{v}^{\pm}_{R,\text{max}}(r_0)\cdot\bm{v}^{\pm}_{R,\text{max}}(r)$ and
$\bm{v}^{\pm}_{R,\text{max}}(r_0)\cdot[\bm{v}^{\pm\pm}_{L,\text{max}}(r)]^{\mathsf{T}}$, are useful to check the spatial dependence of the eigenvector.
Here, we take $r_0=N/4$; $\bm{v}^{p}_{R,L,\text{max}}(r_0)$ is the eigenvector of 
$\hat{\chi}^{p}_{\text{sc}}(N/2+1,N/2+1+N/4)$. See Eq. (\ref{sc_cf}) and the definition 
of $\bm{v}_{\rm sc}^p(r)$ in Sect. \ref{sec_form}. Note that $\bm{v}^{\pm}_{L,\text{max}}(r)$ is a row vector and ``$\mathsf{T}$” represents the transpose.

Let us first discuss the Cooper pairs in the FM phase. Although 
the SC correlations in the FM phase are not enhanced as shown in Fig. \ref{S1Sc}, we here try to explain our analysis used in the following discussions. 
Figures \ref{sc_comp}(a) and \ref{sc_comp}(b) show the $r$ dependences of $|\lambda^{\pm}_{\text{max}}(r)|$ and $\lambda^{\pm}_{\text{max}}(r)$, respectively,
in the FM phase for $h=0.0$, $J=2.0$, $I=1.0$, and $N=120$ with the cutoff $m=500$.
The two eigenvalues $\lambda^{+}_{\text{max}}(r)$ and $\lambda^{-}_{\text{max}}(r)$ are almost the same and exhibit monotonic decreases as $r$ increases.
This means that the SC eigenmodes corresponding to $\lambda^{\pm}_{\text{max}}(r)$ are degenerate.
Now, we discuss which components in Eqs. (\ref{Cooper_in}) and (\ref{Cooper_on}) are dominant for the modes with $\lambda^{\pm}_{\text{max}}(r)$. 
To prevent the notational complexity of Eqs. (\ref{Cooper_in}) and (\ref{Cooper_on}) with the four indices, 
we introduce the following notation for the local operators with the index $n$ that corresponds to the set $(\alpha,p)$:
\begin{align}
	\mathcal{S}^{n}_{j}=\begin{cases}
		\hat{c}_{j,-}\hat{c}_{j,+}{\mathcal O}_j^\alpha,\quad &(n={\rm a,\ b,\ c,\ \cdots}),\\
		\hat{c}_{j,p}{\mathcal O}_j^\alpha,\quad &(n=1,2,3,\cdots).
	\end{cases}\label{eq:LocalOpS}
\end{align}
The list of $\mathcal{S}^{n}_{j}$ is given in Table \ref{list_coop}.
Thus, the Cooper pairs in Eqs. (\ref{Cooper_in}) and (\ref{Cooper_on}) can be represented by 
the product of two $\mathcal{S}^{1,2,\cdots}_{j}$'s for the intersite pairs or a single $\mathcal{S}^{{\rm a,b},\cdots}_{j}$ for the local pairs. We denote them as 
\begin{align}
	&\mathcal{D}^{n,m}_{j,j+1}=\mathcal{S}^{n}_{j}\mathcal{S}^{m}_{j+1}, \quad &(n,m=1,2,3,\cdots),\label{eq:S1}\\
	&\mathcal{D}^{n}_{j,j}=\mathcal{S}^{n}_{j}, \quad &(n={\rm a, b, c,} \cdots).\label{eq:S2}
\end{align}
The weights of the largest nine elements in the amplitude of $\bm{v}^{+}_{R,\text{max}}(r)$ for $r=30$ are shown in Fig. \ref{sc_comp}(c). The two numbers for the horizontal axis correspond to the indices $n$ and $m$ in Eq. (\ref{eq:S1}). 
The two largest components have almost the same magnitude $\sim 0.37$.
In the operator form, they read
\begin{align}
    \mathcal{D}^{6,6}_{j,j+1}=\left(\mathcal{Q}^{-+}_{j}\hat{c}_{j,-}\right)\left(\mathcal{Q}^{-+}_{j+1}\hat{c}_{j+1,-}\right)=\hat{\Delta}^{33}_{--}(j,j+1),\nonumber \\    
    \mathcal{D}^{3,3}_{j,j+1}=\left(\mathcal{Q}^{+-}_{j}\hat{c}_{j,+}\right)\left(\mathcal{Q}^{+-}_{j+1}\hat{c}_{j+1,+}\right)=\hat{\Delta}^{22}_{++}(j,j+1).
\end{align}
Here, $\mathcal{Q}^{pq}_{j}=\ket{p}_{j}\bra{q}_{j}$ represents the local spin operator with $p,q=\pm$ being the spin-inversion parity as defined in Eq. (\ref{eq:Defpm}).
The next six components have almost the same magnitude $\simeq 0.32\sim 0.36$. These eight largest components occupy 97 \% of all the weights.
The same is also true for the eight largest components in $\bm{v}^{-}_{R,\text{max}}(r)$ as shown in Fig. \ref{sc_comp}(d). As shown in Figs. \ref{sc_comp}(e) and \ref{sc_comp}(f), the eigenvectors do not vary as a function of $r$ for $10<r<50$; the inner products $\bm{v}^{\pm}_{\text{R},\text{max}}(30)\cdot\bm{v}^{\pm}_{\text{R},\text{max}}(r)$ and $\bm{v}^{\pm}_{\text{R},\text{max}}(30)\cdot[\bm{v}^{\pm}_{\text{L},\text{max}}(r)]^{\mathsf{T}}$ are almost equal to 1. This indicates that the modes shown in Figs. \ref{sc_comp}(c) and 
\ref{sc_comp}(d) are dominant ones for wide range of $r$.   

Now, let us assume that these leading components have the same magnitude in order to visualize the profile.
After taking the linear combinations of them,
the two degenerate leading Cooper pairs turn out to be
\begin{eqnarray}
	{\mathcal A}_{\downarrow\downarrow}=({\mathcal P}_{j,\uparrow} c_{j,\downarrow}-\hat{S}^+_jc_{j,\uparrow})({\mathcal P}_{j+1,\uparrow} c_{j+1,\downarrow}-\hat{S}^+_{j+1}c_{j+1,\uparrow}),
	\label{coop_fmS1}
\end{eqnarray}
and 
\begin{eqnarray}
	{\mathcal A}_
	{\uparrow\uparrow}=({\mathcal P}_{j,\downarrow} c_{j,\uparrow}-\hat{S}^-_jc_{j,\downarrow})({\mathcal P}_{j+1,\downarrow} c_{j+1,\uparrow}-\hat{S}^-_{j+1}c_{j+1,\downarrow}),
	\label{coop_fmS2}
\end{eqnarray}
where we have introduced the operators ${\mathcal P}_{j,\sigma}=\ket{\sigma}_j\bra{\sigma}_j$ with $\sigma=\uparrow,\downarrow$ and $\hat{S}^{\pm}_{j}=\hat{S}^{x}_{j}\pm {\rm{i}}\hat{S}^{y}_{j}$.
Since the Cooper pairs can be taken to be eigen operators for the $z$-component of the total spin $\hat{S}^{\text{tot}}_z=\sum^{N}_{j=1}(\hat{s}^{z}_{j}+\hat{S}^{z}_{j})$ for $h=0$, we define ${\mathcal A}_{\downarrow\downarrow}({\mathcal A}_{\uparrow\uparrow})$ as ones with the eigenvalue $+1$($-1$). 
Since the electron spins antiferromagnetically couple to the local spins, those representing the local AFM correlations
 are present in Eqs. (\ref{coop_fmS1}) and (\ref{coop_fmS2}). 
 The other combinations such as ${\mathcal P}_{j,\uparrow}c_{j,\uparrow}$, $\hat{S}^+_{j}c_{j,\downarrow}$, etc., represent local FM correlations, and thus, are absent. Note also that the combination 
${\mathcal P}_{j,\downarrow} c_{j,\uparrow}-\hat{S}^-_jc_{j,\downarrow}\sim [\hat{s}_j\cdot \hat{S}_j,c_{j,\uparrow}]\sim -{\rm i}\partial c_{j,\uparrow}/\partial t$\cite{PRB.52.1271},
 where $\partial/\partial t$ is the time $(t)$ derivative, and we have ignored non-composite components as is valid for strong AFM correlation cases. Roughly speaking, ${\mathcal A}_{\sigma\sigma}\sim (\partial c_{j,\sigma}/\partial t )(\partial c_{j+1,\sigma}/\partial t)$.

In the KP phase, the SC correlations are weak.
$\lambda^{+}_{\text{max}}(r)$ is suppressed with $\theta^{+}_{\text{sc}}=2.51$,
and $\lambda^{-}_{\text{max}}(r)$ decays exponentially with $\xi^{-}_{\text{sc}}=0.82$ as shown in Fig. \ref{sc_comp}(g).
Although the SC correlations are weak, the leading SC correlations correspond to the even sector: $\lambda^{+}_{\text{max}}(r)$.
From Fig. \ref{sc_comp}(i), one can obtain the Cooper pairs with the largest amplitude $\simeq \pm 0.5$:
\begin{align}
	\mathcal{D}^{2,3}_{j,j+1},\;\;\mathcal{D}^{3,2}_{j,j+1},\;\;
	\mathcal{D}^{2,2}_{j,j+1},\;\;\mathcal{D}^{3,3}_{j,j+1}.
\end{align}
These four in total correspond to the following Cooper pair,
\begin{align}
	\left(\mathcal{Q}^{+-}_{j}\hat{c}_{j,+}-{\mathcal Q}^{++}_{j}\hat{c}_{j,-}\right)
	\left(\mathcal{Q}^{+-}_{j+1}\hat{c}_{j+1,+}-{\mathcal Q}^{++}_{j+1}\hat{c}_{j+1,-}\right).
	\label{coop_KP}
\end{align}
In the KP phase, the intermediate field $h$ makes the local spins align along the positive $x$-direction and induce the local state $\ket{+}_{j}$,
while the electron spins still couple to the local spins antiferromagnetically owing to the large $J>h$.
Thus, only the operators generating the intermediate state $\ket{+}_{j}$ appear in Eq. (\ref{coop_KP}),
while keeping the local AFM correlations.
$\lambda^{+}_{\text{max}}(r)$ decays without oscillations as shown in Fig. \ref{sc_comp}(h) and the dominant Cooper pairs represented by Eq. (\ref{coop_KP}) are stable over the long distance. See Fig. \ref{sc_comp}(k). In contrast,
$\bm{v}^{-}_{\text{R},\text{max}}(30)\cdot \bm{v}^{-}_{\text{R},\text{max}}(r)$ and $\bm{v}^{-}_{\text{R},\text{max}}(30)\cdot[\bm{v}^{-}_{\text{L},\text{max}}(r)]^{\mathsf{T}}$
show strong oscillations [Fig. \ref{sc_comp}(l)].
Because of them, it is difficult to determine the dominant mode for the odd sector. Note that 
Fig. \ref{sc_comp}(j) represents the eigenvector for $r=30$, which differs from that for sufficiently large $r$. Fortunately, they are not the dominant ones, and thus, not important.

\begin{figure}[t!]
	\centering
	\includegraphics[width=0.9\linewidth]{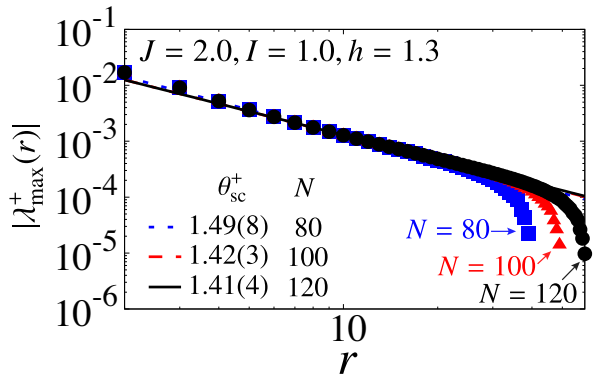}	
	\caption{(Color online)
	$r$ dependence of $|\lambda^{+}_{\text{max}}(r)|$ for $I=1.0$, $J=2.0$, $h=1.3$, $n_c=1/2$, $N=80$, $100$ and $120$ with the cutoff $m=700$.
	}
	\label{sc_pow_size}
\end{figure}

When $J=2.0$ and $h=1.3$, in the TLL phase,
the SC correlations in the even parity sector are strongly enhanced 
with $\theta^{+}_{\text{sc}}=1.41(4)$ as depicted in Figs. \ref{sc_comp}(m) and \ref{sc_comp}(n).
To confirm that this enhancement occurs in the thermodynamic limit,
we calculate $\theta^{+}_{\text{sc}}$ for $N=80$, $100$, and $120$ with the cutoff $m=700$. The results are shown in Fig. \ref{sc_pow_size} and 
 indicate that $\theta^{+}_{\text{sc}}$ tends to be slightly smaller with increasing $N$. From this, we conclude that $N=120$ is sufficient to study the bulk properties.
The Cooper pairs corresponding to the enhanced SC correlations are read from Fig. \ref{sc_comp}(o):
\begin{align}
	\mathcal{D}^{2,2}_{j,j+1}=\left(\mathcal{Q}^{++}_{j}\hat{c}_{j,-}\right)\left(\mathcal{Q}^{++}_{j+1}\hat{c}_{j+1,-}\right)=\hat{\Delta}^{11}_{--}(j,j+1),\label{Delta22}\\
	\mathcal{D}^{1,1}_{j,j+1}=\left(\mathcal{Q}^{++}_{j}\hat{c}_{j,+}\right)\left(\mathcal{Q}^{++}_{j+1}\hat{c}_{j+1,+}\right)=\hat{\Delta}^{11}_{++}(j,j+1).\label{Delta11}
\end{align}
$\mathcal{D}^{2,2}_{j,j+1}$ and $\mathcal{D}^{1,1}_{j,j+1}$ have the amplitude $\simeq0.6$ and $\simeq0.5$, respectively, and they occupy 61 \% among all the weights.
The appearance of these Cooper pairs is qualitatively understood as follows.
Since the local spins almost aligned along the positive $x$-direction by the large $h$,
the conduction electrons can couple to the local spins only through $\mathcal{Q}^{++}_{j}=\ket{+}_{j}\bra{+}_{j}$, otherwise vanishes if acting on the ground state. 
Since the KP phase exists nearby, the AFM correlations represented by $\ket{+}_{j}\bra{+}_{j}\hat{c}_{j,-}$ still exist in Eq. (\ref{Delta22}). The clear distinction from the Cooper pairs in the FM and the KP phases lies in the fact that the second leading pair represented by Eq. (\ref{Delta11}) consists of the ferromagnetically coupled one with $\ket{+}_{j}\bra{+}_{j}\hat{c}_{j,+}$. This strongly suggests that the SC correlations are enhanced by the spin fluctuations.

Assuming that these Cooper pairs have the same amplitude,
the leading Composite Cooper pair is represented by
\begin{align}
	{\mathcal A}=\left(\hat{c}_{j,-}\hat{c}_{j+1,-}+\hat{c}_{j,+}\hat{c}_{j+1,+}\right){\mathcal Q}^{++}_{j}{\mathcal Q}^{++}_{j+1}.
	\label{coop_TLL}
\end{align}
This form of the composite Cooper pair with even spin-inversion parity is another main result in this paper.
${\mathcal A}$ can be regarded as the equal-spin pairing between the nearest neighbors under the background of polarized local moments. \tred{ Note that this composite pair includes $cc$-only conventional component with $\sim 15$\%. This is easily estimated from the definition of $\mathcal Q^{++}$.
We have also calculated the conventional Cooper-pair correlation functions. The results show the same exponent $\theta_{\rm sc}^+$ as that for Eq. (\ref{coop_TLL}) but with smaller amplitude.
}
We have checked that qualitatively the same Cooper pairs are formed and their correlations are enhanced for $n_c=3/10$ and $4/5$.
It is evident from Fig. \ref{sc_comp}(q) that this Cooper pair in Eq. (\ref{coop_TLL}) is stable  over the wide range of the system.
We also show the Cooper pair components in the odd sector in Fig. \ref{sc_comp}(p) and its stability in Fig. \ref{sc_comp}(r).
The inner products $\bm{v}^{-}_{\text{R},\text{max}}(30)\cdot\bm{v}^{-}_{\text{R},\text{max}}(r)$ and $\bm{v}^{-}_{\text{R},\text{max}}(30)\cdot[\bm{v}^{-}_{\text{L},\text{max}}(r)]^{\mathsf T}$ show oscillations, which suggests the spatially modulated superconductivity.
However, the corresponding SC correlations are not enhanced $\theta^{-}_{\rm sc}>2$ as shown in Fig. \ref{sc_comp}(m) and they are not important.

Finally, we explain the SC correlations for $h=3.0$ in the TLL phase close to the FP phase. 
As discussed in Sect. \ref{sec_pow}, near the saturation field, the value of $\theta^{-}_{\rm sc}$ is underestimated. This underestimation arises from the presence of 
the small Fermi wavenumber $k^{\text{up}}_{\text{F}}$ [Eq. (\ref{kf})].
We calculate the SC correlation matrix in the odd sector [Eq. (\ref{sc_cf})] 
for the free fermion case with $k^{\text{up}}_{\text{F}}=2\pi/N$.
As is clearly shown in Figs. \ref{sc_comp}(s) and \ref{sc_comp}(t), the $r$ dependence of $|\lambda^{-}_{\text{max}}(r)|$ almost coincides with that for the free fermion case. 
The oscillations in $\bm{v}^{-}_{\text{R},\text{max}}(30)\cdot\bm{v}^{-}_{\text{R},\text{max}}(r)$ and $\bm{v}^{-}_{\text{R},\text{max}}(30)\cdot[\bm{v}^{-}_{\text{L},\text{max}}(r)]^{\mathsf T}$[Fig. \ref{sc_comp}(x)] also originate from the large and small Fermi wavenumbers $k^{\text{low}}_{\text{F}}$ and $k^{\text{up}}_{\text{F}}$.
Thus, we conclude that $\theta^{-}_{\text{sc}}\sim 1.2$ does not represent the enhancement of the SC correlations.

One can obtain the Cooper pair eigenvector in the odd sector from Fig. \ref{sc_comp}(v),
which is qualitatively similar to that for $h=1.3$ [Fig. \ref{sc_comp}(p)].
For $h\sim 3$ or higher, the local and electron spins are almost polarized along the positive $x$-direction.
This leads to the appearance of the Cooper pair of the form of the second term in Eq. (\ref{coop_TLL}), in the even sector for the wide range of the system [Figs. \ref{sc_comp}(u) and \ref{sc_comp}(w)],
but the SC correlation itself is not enhanced.

\subsection{Results for the spin-1 KLM}\label{secKLM1}
In this subsection, we explain the SC correlations for the spin-1 KLM.
Unlike the case of the spin-1/2 KLM, there is no sign of the enhancement of the SC correlations,
namely, $\theta^{\pm}_{\text{sc}}\gtrsim 2$ or $\lambda^{\pm}_{\text{max}}(r)$ decays exponentially for all the phases.

\begin{figure}[t]
	\centering
	\includegraphics[width=0.9\linewidth]{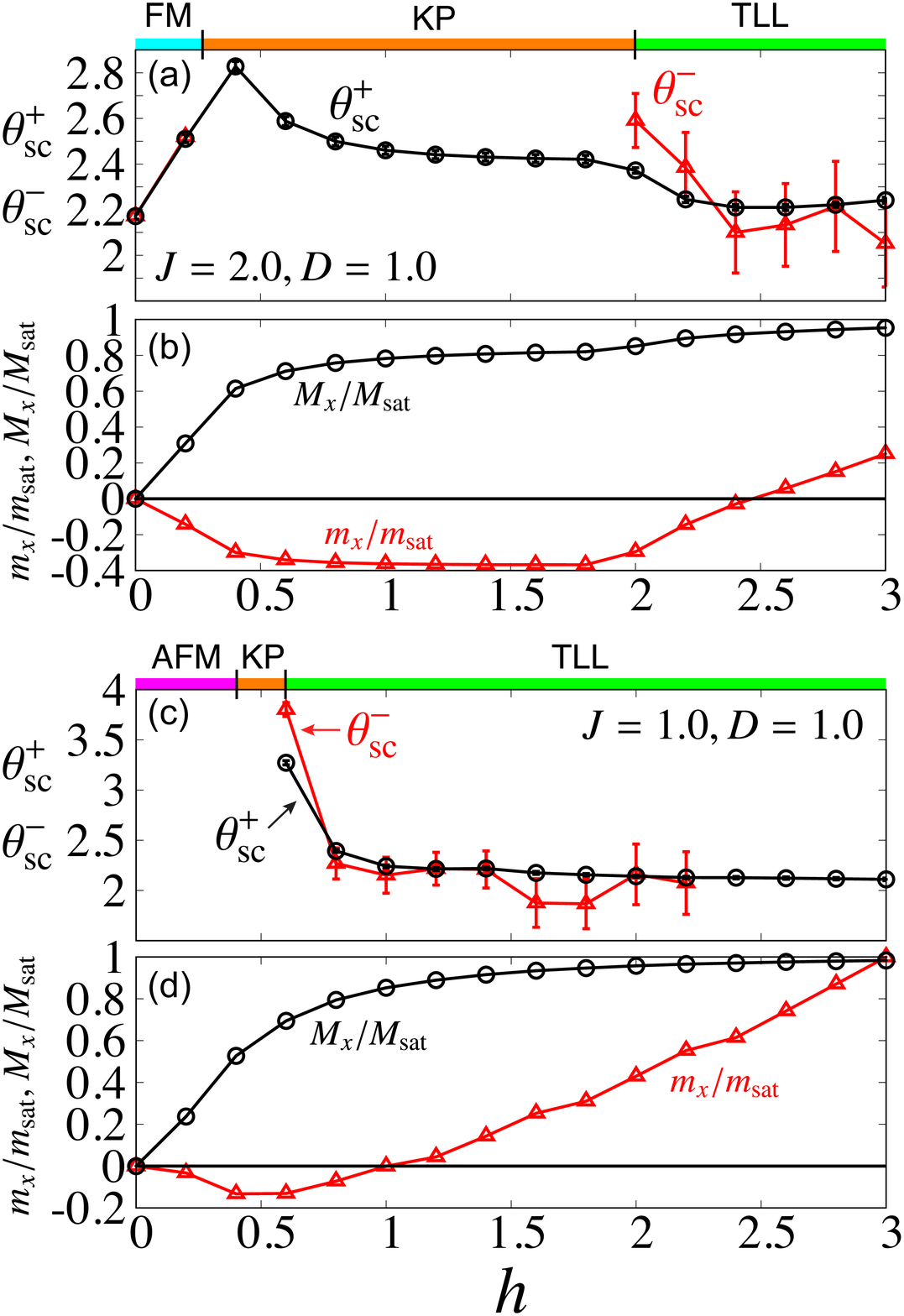}	
	\caption{(Color online)
	$h$ dependences of the exponent $\theta^{\pm}_{\text{sc}}$ and the magnetizations $m_x$ and $M_x$ along the $x$-direction for the spin-1 KLM.
	The calculations have been carried out for $D=1.0$, $n_c=1/2$, $N=120$ with the cutoff $m=400$.
	(a) and (b) $J=2.0$, (c) and (d) $J=1.0$.
	$\theta^{\pm}_{\text{sc}}$ cannot be defined for the AFM phase and $\theta^{-}_{\text{sc}}$ for the KP phase.
	}
	\label{S2Sc}
\end{figure}

\begin{figure}[t]
	\centering
	\includegraphics[width=0.9\linewidth]{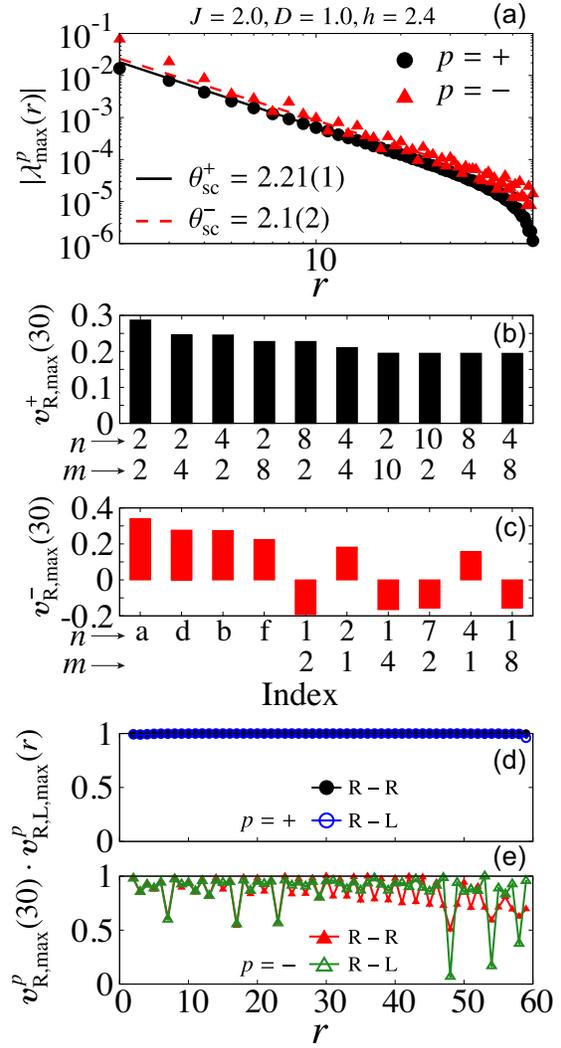}	
	\caption{(Color online)
	Distance $r$ dependences of (a) $|\lambda^{\pm}_{\text{max}}(r)|$ and (b), (c) the elements of the corresponding eigenvectors $\bm{v}^{\pm}_{\text{R,max}}(r=30)$ for $J=2.0$, $D=1.0$, $h=2.4$, $n_c=1/2$, and $N=120$ with the cutoff $m=400$ in the TLL phase for the spin-1 KLM.
	(d) and (e) $r$ dependences of the inner product $\bm{v}^{\pm}_{\text{R,max}}(30)\cdot\bm{v}^{\pm}_{\text{R,max}}(r)$ and $\bm{v}^{\pm}_{\text{R,max}}(30)\cdot[\bm{v}^{\pm}_{\text{L,max}}(r)]^\mathsf{T}$.
	}
	\label{S2Sc_comp}
\end{figure}

Figure \ref{S2Sc} shows the $h$ dependences of $\theta^{\pm}_{\text{sc}}$ and the magnetizations for $N=120$, $n_c=1/2$, $D=1.0$ and $J=1.0$ and $2.0$ with the cutoff $m=400$.
$\lambda^{+}_{\text{max}}(r)$ shows a power-law decay for the FM, KP, and TLL phases,
while $\lambda^{-}_{\text{max}}(r)$ decays exponentially for the KP phase, and thus, not shown.
These tendencies are similar to the case of the spin-1/2 KLM.
The differences between the results of the spin-1/2 and spin-1 KLMs are as follows.
First, the SC correlations show no enhancement in all the region of the magnetic fields; $\theta^{\pm}_{\rm sc}>2$. 
The detailed behavior is also different. For example, $\theta^+_{\rm sc}$ are suppressed around the phase boundary between the FM and KP phases as shown in Fig. \ref{S2Sc}(a), while for the spin-1/2 case, the SC correlations are suppressed in the middle of the KP phase. See Fig. \ref{S1Sc}(c).
Second, in the TLL phase, the values of $\theta^-_{\rm sc}$ and $\theta^+_{\rm sc}$ are similar as shown in Figs. \ref{S2Sc}(a) and \ref{S2Sc}(c).
The SC correlations themselves are larger for the odd-parity sector than those for the even parity sector [$|\lambda^{-}_{\text{max}}(r)|>|\lambda^{+}_{\text{max}}(r)|$, as shown in Fig. \ref{S2Sc_comp}(a)]. 
Note that for the spin-1/2 case, the SC correlations for the even parity are most enhanced as discussed in Fig. \ref{S1Sc}(c).
For $J=1.0$, there is an AFM phase for small $h$.
 We find that the SC correlations are strongly suppressed in the AFM phase,
where $\lambda^{\pm}_{\text{max}}(r)$ shows an exponential decay and $\theta^{\pm}_{\text{sc}}$ is not shown in Fig. \ref{S2Sc}(c).

As discussed in Sect. \ref{sec_pow}, the estimation of $\theta^{-}_{\text{sc}}$ in the TLL phase close to the FP phase is difficult owing to the presence of the long wavelength oscillations with $\lambda^{\text{up}}_{\text{F}}$.
Actually, the fitting of $\theta^{-}_{\text{sc}}$ does not work for $J=1.0$ and $h\geq 2.4$ [Fig. \ref{S2Sc}(c)].

Figures \ref{S2Sc}(b) and \ref{S2Sc}(d) show the $h$ dependences of the magnetizations $m_x/m_{\text{sat}}$ and $M_x/M_{\text{sat}}$ along the $x$-direction for $J=2.0$ and $J=1.0$, respectively.
Note that the saturation values of the magnetizations $m_{\text{sat}}=n_c/2=0.25$ and $M_{\text{sat}}=1.0$.
For the spin-1/2 KLM (Fig. \ref{S1Sc}), we have pointed out the enhancement of the SC correlations occurs when $m_x\simeq 0$.
However, for the spin-1 KLM, there is no sign of the clear increases in the SC correlations even when $m_x\simeq 0$ for $(J,h)=(2.0,2.4)$ and $(1.0,1.0)$.

Now, we briefly discuss the $r$ dependence of $\lambda^{\pm}_{\text{max}}(r)$ and 
the corresponding SC correlations in the TLL phase.
Figure \ref{S2Sc_comp} shows the profile of the SC correlations  
for $N=120$, $n_c=1/2$, $D=1.0$, $J=2.0$, and $h=2.4$ with the cutoff $m=400$. For these parameters, 
 $m_x/m_{\text{sat}}\simeq 0$ and the value of $\theta^{+}_{\text{sc}}$ takes the local minimum $\simeq 2.21(1)$ as a function of $h$.
One can see $\lambda^{-}_{\text{max}}(r)>\lambda^{+}_{\text{max}}(r)$ in Fig. \ref{S2Sc_comp}(a),
and this means that the SC correlations in the odd sector are dominant in the TLL phase.
The weights of the largest ten elements in $\bm{v}^{\pm}_{\text{R},\text{max}}(30)$ are shown in Figs. \ref{S2Sc_comp}(b) and \ref{S2Sc_comp}(c).
One can read the corresponding Cooper pairs from Table \ref{list_coop} as discussed in Sect. \ref{sec_coop}.
Unfortunately, it is difficult to deduce the intuitive form from the distribution in Figs. \ref{S2Sc_comp}(b) and \ref{S2Sc_comp}(c), since many elements have similar weights.
In addition to this, the inner product $\bm{v}^{-}_{\text{R},\text{max}}(30)\cdot\bm{v}^{-}_{\text{R},\text{max}}(r)$ and 
$\bm{v}^{-}_{\text{R},\text{max}}(30)\cdot[\bm{v}^{-}_{\text{L},\text{max}}(r)]^\mathsf{T}$
shows oscillations [Fig. \ref{S2Sc_comp}(e)],
which suggests that the mode is spatially modulated and also causes the difficulty in identifying the dominant Cooper pairs in the odd sector. In contrast, the Cooper pairs in the even sector are stable over the long distance,
since $\bm{v}^{+}_{\text{R},\text{max}}(30)\cdot\bm{v}^{+}_{\text{R},\text{max}}(r)$ and 
$\bm{v}^{+}_{\text{R},\text{max}}(30)\cdot[\bm{v}^{+}_{\text{L},\text{max}}(r)]^\mathsf{T}$ takes the value $\simeq 1$ [Fig. \ref{S2Sc_comp}(d)].
However, the SC correlations corresponding to $\bm{v}^{+}_{\text{R},\text{max}}(r)$ are not the dominant ones.

Although there remain uncertainties in the determinations of the dominant Cooper pairs,
we conclude that the SC correlations are not enhanced for the spin-1 KLM.
Since there are a finite energy cost $\simeq D$ to flip the local spin $\ket{\Uparrow},\ket{\Downarrow}\rightarrow\ket{\mathbb O}$ for generating spin-fluctuations $\ket{\Uparrow}\leftrightarrow\ket{\Downarrow}$, while in the spin-1/2 KLM, the flip of the local spins are possible without energy costs as long as the single-ion potential is concerned.
The presence of the intermediate state $\ket{\mathbb O}$ is expected to be one of the main reasons why the spin-1 KLM shows no enhancement of the SC correlations.

\section{Discussions}\label{sec_dis}
In this section, we will discuss our results and compare them to the experimental data of the FM superconductor URhGe.
In Sect. \ref{sec_mag_sz}, we will briefly discuss the effects of the longitudinal magnetic field. 
Unlike the case of the transverse magnetic field,
the longitudinal magnetic field does not enhance the SC correlations.
In Sect. \ref{sec_diff}, we will summarize the difference between the spin-1/2 and spin-1 KLMs.
\tblue{Finally,} in Sect. \ref{sec_dis_compare}, we will compare our results with the experimental data and those in other theoretical studies. \tred{The limit of the strong Kondo coupling is examined. It turns out that the SC correlations are enhanced at zero magnetic field inside the FM phase. }

\subsection{Effects of longitudinal magnetic fields}\label{sec_mag_sz}

So far, we have discussed the situation where the magnetic field is applied to the transverse direction. One might wonder what happens for the longitudinal fields?  
In this subsection, we briefly discuss the SC correlations under the longitudinal magnetic fields $h_z$ for the spin-1/2 KLM.
The magnetic field term in Eq. (\ref{S1_HAM}) is replaced by $-h_z\sum^{N}_{j=1}(\hat{s}^{z}_{j}+\hat{S}^{z}_{j})$.
We calculate the SC correlations in the same manner as in the case of the transverse magnetic field,
and we find no clear enhancement of the SC correlations.

As discussed in Sect. \ref{sec_coop} [see the line of discussion for Eqs. (\ref{coop_fmS1}) and (\ref{coop_fmS2})], 
all the Cooper pairs can be taken to be eigen operators for $\hat{S}^{\text{tot}}_z=\sum^{N}_{j=1}(\hat{s}^{z}_{j}+\hat{S}^{z}_{j})$ owing to the presence of $U(1)$ symmetry for the $z$-component of the spins.
Thus, the SC correlation matrix [Eq. (\ref{sc_cf})] can be characterized by the Cooper pair's $z$-component of the spin: $s_z=\pm 3,\pm 2,\pm 1,0$.
We define the exponents similarly to Eq. (\ref{sc_pow}):
\begin{align}
	|\lambda^{(s_z)}_{\text{max}}(r)|\sim r^{-\theta^{(s_z)}_{\text{sc}}},\quad (s_z=\pm 3,\pm 2,\pm 1,0).
\end{align}
Here, $\lambda^{(s_z)}_{\text{max}}(r)$ corresponds to the dominant Cooper pairs with the eigenvalue $s_z$.
We also define the average value of the Cooper pairs:
\begin{align}
	\Delta^{(s_z)}_{\text{av}}=\sum^{20}_{r=5}\sqrt{|\lambda^{(s_z)}_{\text{max}}(r)|},
\end{align}
which can also characterize the enhancement of the SC correlations.

\begin{figure}[t]
	\centering
	\includegraphics[width=\linewidth]{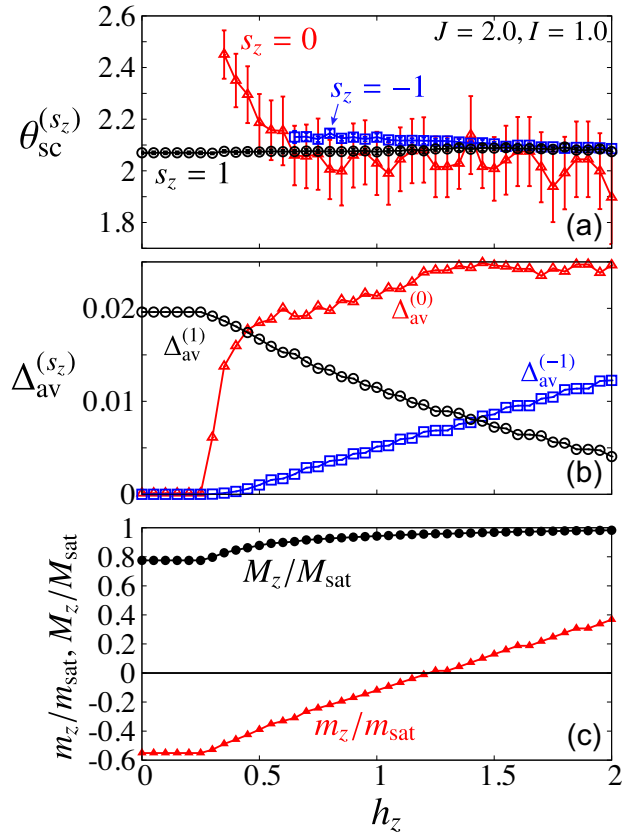}	
	\caption{(Color online)
	$h_z$ dependences of (a) $\theta^{(s_z)}_{\text{sc}}$, (b) $\Delta^{(s_z)}_{\text{av}}$, and (c) the magnetizations $m_z$ and $M_z$ along the $z$-direction for $I=1.0$, $J=2.0$, $n_c=1/2$, and $N=120$ with the cutoff $m=600$.
	The magnetic field $h_z$ is applied along the $z$-direction.
	}
	\label{S1sc_z}
\end{figure}

Figure \ref{S1sc_z} shows the results for \tred{$N=120$, $n_c=1/2$, $J=2.0$ and $I=1.0$ with the cutoff $m=600$.}
Note that the ground state is always degenerate with $\pm S^{\text{tot}}_z$.
We will show the data for $S^{\text{tot}}_z\geq 0$ sectors.
We find that $\lambda^{(s_z)}_{\text{max}}(r)$ with $s_z=-1,0,1$ shows a power-law decay as shown in Fig. \ref{S1sc_z}(a),
and the other SC correlations decay exponentially (not shown).
In particular, the SC correlations with $s_z=1$ are dominant in the FM phase for $h_z\lesssim 0.25$,
while those with $s_z=0$ are dominant in the paramagnetic phase $h_z\gtrsim 0.25$.
Note that the dominant Cooper pair in the FM phase is ${\mathcal A}_{\downarrow\downarrow}$ in Eq. (\ref{coop_fmS1}); ``$\downarrow\downarrow$'' owing to the fact that the definition Eq. (\ref{coop_fmS1}) is based on the annihilation operators. We have checked that 
${\mathcal A}_{\uparrow\uparrow}$ in Eq. (\ref{coop_fmS2}) is dominant for the other 
FM domain with $S_z^{\rm tot}<0$. 
We also present $h_z$ dependence of $\Delta^{(s_z)}_{\text{av}}$ in Fig. \ref{S1sc_z}(b).
One can clearly see that the SC correlations for $s_z=1$ are largest in the FM phase and 
those for $s_z=0$ are suddenly induced around $h\simeq 0.25$,
where the FM phase is destabilized by $h_z$ and the magnetizations along the $z$-direction start to increase as shown in Fig. \ref{S1sc_z}(c).
The SC correlations with $s_z=-1$ are gradually induced by $h_z$ toward the FP phase.

Among all $\lambda^{(s_z)}_{\text{max}}(r)$, the SC correlations with $s_z=1$ are dominant in the FM phase
and those for $s_z=0$ are induced by $h_z$.
However, the exponents $\theta^{(0)}_{\text{sc}}$ and $\theta^{(1)}_{\text{sc}}$ are always larger than $\theta^{\text{free}}_{\text{sc}}=2$ or $\sim 2$ within the error bars as depicted in Fig. \ref{S1sc_z}(a).
Thus, we conclude that the SC correlations shows no noticeable enhancement under the longitudinal magnetic field in a stark contrast to the case of the transverse field in the Ising anisotropic spin-1/2 KLM.

\subsection{Spin-1/2 vs spin-1}\label{sec_diff}
In this paper, we have studied the SC correlations in the spin-1/2 and spin-1 KLMs under the magnetic fields.
The enhancement of the SC correlations occurs in the spin-1/2 KLM, when the magnetic field is applied along the transverse direction ($x$-direction).
In Sect. \ref{sec_sc}, we have discussed this, paying our attention to the vanishing conduction electron magnetization $m_x\sim 0$ near the KP phase. See Fig. \ref{S1Sc}.
This is a kind of the Jaccarino--Peter effects\cite{PRL.9.290},
although the present Cooper pairs are composite ones and equal-spin pairing if focusing on the conduction electron part.  
When $m_x\sim 0$, the effective magnetic field felt by the conduction electrons is small.
It is natural to expect that there is large spin-fluctuation,
which is one of the relevant mechanisms for this enhanced SC correlation.

This view can also explain why there is no noticeable enhancement of the SC correlations in the spin-1 KLM as discussed in Sect. \ref{secKLM1}.
The single-ion anisotropy strongly suppresses the local spin-flip, leading to suppression of the conduction electron spin fluctuations.
Note that this reasoning is valid only when the exchange interactions contain terms that change the spin $\pm 1$.
In general, larger local spins have many intermediate states between the polarized states $\ket{S}$ and $\ket{-S}$,
and the presence of these states suppresses the fluctuations of the local and electron spins.

The above argument, of course, just represents one aspect. One can consider other origins for the mechanism of enhanced SC correlations. Indeed, we have calculated the SC correlations in the spin-1 KLM with Ising interactions. Namely, the Hamiltonian is the same as in Eq. (\ref{S1_HAM}) except for the magnitude of the local spins. 
This model shows the enhancement of the SC correlations $\theta^{+}_{\text{sc}}\simeq1.71(2)$ for $J=1.5$, $h=1.4$, and $n_c=1/2$ in the TLL phase. 
Thus, in addition to the magnitude of the local spins, the types of the Ising anisotropy are also important for the SC correlations.
It seems that superconductivity favors the exchange anisotropy rather than the single-ion anisotropy.

\subsection{Comparison with the experiments and other theories}
\label{sec_dis_compare}
As mentioned in Introduction, URhGe shows reentrant superconductivity at the magnetic field $H_b\sim 8$ T\cite{URhGe,URhGe_RSC}
 and this magnetic field corresponds to the phase boundary between the FM and paramagnetic phases under the field.
There is also a metamagnetic increase as increasing the magnetic field at around 8 T\cite{URhGe_RSC,PRB.83.195107}.
One of the authors has discussed the reentrant superconductivity with the mean-field and spin-wave approximations\cite{PRB.87.064501}.
The weak point there is that the mean-field treatment cannot take into account the metamagnetic fluctuations.
In this respect, the present analysis can fully take into account such fluctuations,
and the results indeed show both enhanced SC correlations and metamagnetic behavior.
For example, the magnetization curve in Fig. \ref{S1Sc}(e) is qualitatively similar to the experimental data.\cite{PRB.83.195107}
The SC correlations are enhanced [Fig. \ref{S1Sc}(b)] near the magnetic field where the metamagnetic increase occurs.
This is at the FM-KP boundary, but the KP phase is very narrow in this parameter.
Although it is very hard to translate the TLL to the real three-dimensional systems, the physics behind the competition among the transverse field, the Ising anisotropy, and the Kondo correlation is expected to be common beyond the one-dimensional systems.

\begin{figure}[t]
	\centering
	\includegraphics[width=\linewidth]{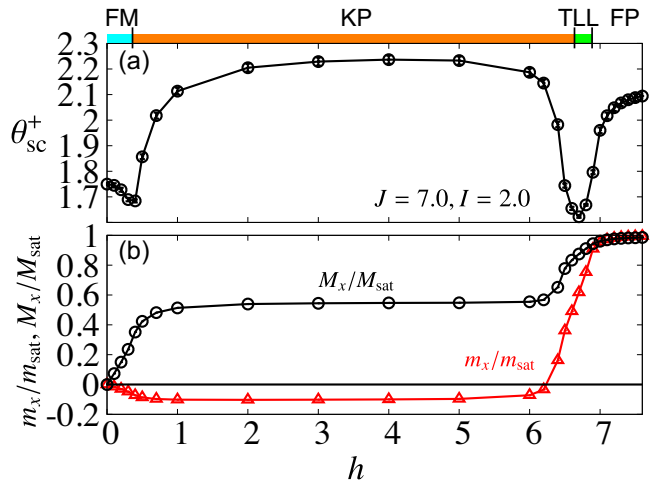}	
	\caption{(Color online)
	$h$ dependences of (a) $\theta^{+}_{\text{sc}}$ and (b) $m_x$ and $M_x$ for $I=2.0$, $J=7.0$, $n_c=1/2$, and $N=120$ with the cutoff $m=500$. }
	\label{fig-largeJlimit}
\end{figure}

We must note that the results shown so far do not explain the superconductivity inside the FM states.
\tred{At zero magnetic field, the exponent $\theta_{\rm sc}^\pm$ for the SC correlations is larger than 2 as shown in Figs. \ref{S1Sc} and \ref{S1sc_z}. 
However, we find that the exponent decreases as increasing the Kondo coupling $J$ and/or the intersite Ising interaction $I$.
For $J\gg t$ with $J>I$, we obtain $\theta_{\rm sc}^+<2$ inside the FM phase. 
Figure \ref{fig-largeJlimit}(a) shows the $h$ dependence of $\theta^{+}_{\text{sc}}$ for $I=2.0$, $J=7.0$, and $n_c=1/2$.
For this large $J$ limit, the SC correlations are highly enhanced even at $h=0$ inside the FM phase, and those in the TLL phase are also enhanced. The Cooper pair eigenvector in the low-field regime is similar to Eqs. (\ref{coop_fmS1}) and (\ref{coop_fmS2}). 
Note that owing to the large $J$, the KP phase is very wide, while the TLL state shrinks compared with the cases for the smaller $J$. See also Fig. \ref{fig-largeJlimit}(b) for the magnetization plateau.}

\tred{
This tendency can be understood by analyzing the effective interaction for the large $J$ limit.\cite{RMP.69.809} 
For large $J$, FM  correlations are developed. The direct intersite FM interaction $I$ enhances such correlations,
leading to the enhancement of the attractive interactions among the Kondo singlets for large $J$. Note that $\theta^+_{\rm sc}>2$ for $I=0$, $J=7.0$, and $n_c=1/2$.  
This is similar to the situation in the $t$--$J$ model\cite{PhysRevB.83.205113}. In the $t$--$J$ model, there is a phase separation for large $J$, while for the parameters in Fig. \ref{fig-largeJlimit}, there is no phase separation.
 Although the energy scale is not compatible with the realistic situation,
it is interesting that these two different SC states are present for large $J$.
This reminds us of the experimental findings in URhGe.\cite{URhGe,URhGe_RSC,PRL.114.216401,PRB.93.201112} Considering the mass renormalization owing to the correlation effects,
one can expect such strong-coupling parameters in the effective model if the parameters are fine-tuned, although the microscopic reasoning is unclear.}

\tred{In the bosonization analysis of the one-dimensional anisotropic KLM,\cite{PhysRevLett.77.1342} it was pointed out that composite triplet pairing correlations are critical at  zero magnetic field. Since this is realized in AFM state with a finite spin gap, it is interesting to explore superconductivity coexisting the ferromagnetic state 
in different model settings.}
To construct microscopic models in which FM superconductivity emerges is a nontrivial challenge and this remains as one of our future works.
	
\tred{ Now, let us discuss the SC correlations in the TLL phase, i.e., in large transverse fields.} 	
In our calculations, the SC correlations for the smaller $J$ region continuously change to those for the larger $J$ in the phase diagram in Fig. \ref{sc_phase}. 
Note that for large $J$, there is a field at which $m_x=0$ and the composite pairs are approximately in the form of Eq. (\ref{coop_TLL}).
For smaller $J$, however, $m_x$ is always positive as shown in Fig. \ref{S1Sc}(e). Then, an imbalance between Eqs. (\ref{Delta22}) and (\ref{Delta11}) is generally induced.
As discussed in Sect. \ref{sec_diff}, this is a kind of the Jaccarino--Peter superconductivity but is different from the zero-spin pairing state owing to the Jaccarino--Peter effect appearing in the Ising-magnon mechanism\cite{PRB.87.064501}.
The composite pairs (\ref{coop_TLL}) are rather similar to the equal spin pairing state near the saturation field\cite{PRB.87.064501} except for the ``compositeness''.
Our results indicate that both types of the SC correlations continuously crossover with each other and the physics behind them is common. 
	
For quantitative comparison between the theory and experiments, further effort is needed.
For this, it is necessary to take into account the realistic band structure including the band touchings\cite{PRL.121.097001},
the microscopic interactions, and the orbital-pair breaking effect\cite{PRB.87.064501} that is absent in one-dimensional models.

\section{Summary}\label{sec_sum}
We have analyzed superconducting correlations in the spin-1/2 and spin-1 Kondo 
lattice models by using the density matrix renormalization group.
Our main result is that superconducting correlations for the spin-1/2 model 
are strongly enhanced under transverse magnetic fields.
\tred{The form of the Cooper pairs has been analyzed and the Cooper pairs and the local spins are 
strongly correlated, which results in the mixture of conventional and composite Cooper pairs in the leading superconducting correlations.}
The condition for the enhancement is related to the zero magnetization of 
the conduction electrons, where strong fluctuations are expected.
This condition is related to the competitions between the Kondo-singlet 
formation and transverse magnetic fields.
\tred{This Cooper pairs are distinct with those in the smaller fields realized in the FM phase for the large $J$ limit as discussed in Sect. \ref{sec_dis_compare}.
} For the spin-1 model, we have not found enhancement of the SC correlations, possibly owing to the fact that the single-ion anisotropy prevents the local spins from fluctuating.
We have also checked there is no clear enhancement of the SC correlations under longitudinal magnetic fields. 
Our results give a possible explanation for the superconductivity under the transverse field in URhGe and related compounds.
For larger spin models, our results suggest that there need direct-fluctuation path that can skip intermediate state with high energy.
This depends microscopic models and materials themselves.
To construct and/or understand the microscopic model for real materials is our future work.

\section*{Acknowledge}
The authors thank K. Ueda and H. Kusunose for fruitful discussions. 
This work was supported by a Grant-in-Aid for Scientific Research (Grant Nos. 16H01079, 16H04017, and 18K03522) from the Japan Society for the Promotion of Science.

\newpage
\appendix
\section{Spin-inversion parity bases}\label{append}
In this Appendix, we summarize the notation related to the spin-inversion parity bases.
The spin-inversion operation in the spin-1/2 KLM is
\begin{equation}
	\ket{\uparrow}_{j} \leftrightarrow \ket{\downarrow}_{j}, \quad {\rm and}\quad 
	\hat{c}_{j,\uparrow}\leftrightarrow \hat{c}_{j,\downarrow},
\end{equation} 
while for the spin-1 KLM,
\begin{equation}
	\ket{\Uparrow}_{j} \leftrightarrow \ket{\Downarrow}_{j}, \quad {\rm and}\quad 
	\hat{c}_{j,\uparrow}\leftrightarrow \hat{c}_{j,\downarrow},
\end{equation} 
keeping $|{\mathbb O}\rangle_j$ unchanged. Under these transformations, 
the Hamiltonians Eqs. (\ref{S1_HAM}) and (\ref{S2_HAM}) are invariant. 
Since this is a kind of parity operation, the eigenvalues for the spin-inversion parity $P$ 
should be $P=1$ (even) or $-1$ (odd).

As defined in the main text, for the spin-1/2 KLM, the spin-inversion parity bases for the local spin $|p \rangle_j$ and the conduction electrons $\hat{c}_{j,p}$ are given by Eq. (\ref{eq:Defpm}) and Eq. (\ref{eq:Defpm2}), respectively, where $p=\pm$ represent the spin-inversion parity index. 
We note that $p=+(-)$ coincides with the spin up (down) component along the $x$-direction. 

Now, we list the spin-inversion parity bases for the spin-1 KLM, for completeness.  
Since the conduction electrons are the same, the difference is in the local spin part and 
there are two kinds of even parity states:
\begin{eqnarray}
	\ket{+}_{j} &\equiv& \frac{1}{\sqrt{2}} \left(\ket{\Uparrow}_{j} + \ket{\Downarrow}_{j}\right),\\
	\ket{0}_{j} &\equiv& \ket{\mathbb O}_{j},
\end{eqnarray}
while the odd-parity state is given by
\begin{eqnarray}
	\ket{-}_{j} &\equiv& \frac{1}{\sqrt{2}} \left(\ket{\Uparrow}_{j} - \ket{\Downarrow}_{j}\right).
\end{eqnarray}
As an important difference from the spin-1/2 KLM, $|\pm\rangle_j$ is not the eigenstates of $\hat{S}^x_j$, since
\begin{align}
	\ket{\Rightarrow}_{j}&=\frac{\ket{\Uparrow}_{j}+\sqrt{2}\ket{\mathbb O}_{j}+\ket{\Downarrow}_{j}}{2},
	\quad {\rm with}\quad\hat{S}^x_j \ket{\Rightarrow}_{j}=+\ket{\Rightarrow}_{j},\nonumber\\
	\ket{\Leftarrow}_{j}&=\frac{\ket{\Uparrow}_{j}-\sqrt{2}\ket{\mathbb O}_{j}+\ket{\Downarrow}_{j}}{2},
	\quad {\rm with}\quad\hat{S}^x_j \ket{\Leftarrow}_{j}=-\ket{\Leftarrow}_{j}.
\end{align}

\section{Definition of Composite Cooper Pair Components 2}\label{append2}
In this Appendix, we summarize the basis operators ${\mathcal S}^n$ ($n=$a, b, c,$\cdots,1,2,3,\cdots$) 
for representing the Cooper pair amplitude, which have been introduced in Eq. (\ref{eq:LocalOpS}). 
 For the spin-1/2 (spin-1) KLM, $n=$ a-d (a-i) for those consisting of two annihilation operators, while $n=1-8$ ($1-18$) for single-annihilation operators as listed in Table \ref{list_coop}. Note that ${\mathcal S}^n$ has a definite spin-inversion parity, which is indicated by $p$ in Table \ref{list_coop}. The spin-inversion parity indices are useful for classifying 
various components of the Cooper pairs in Eqs. (\ref{eq:S1}) and (\ref{eq:S2}).

\begin{table}[t]
	\caption{Definitions of the local operators $\mathcal{S}^n_j=\hat{c}_{j,-}\hat{c}_{j,+}{\mathcal O}^\alpha_j$, $\hat{c}_{j,p}{\mathcal O}^\alpha_j$ at the site $j$ for the spin-1/2 KLM and the spin-1 KLM. $\alpha$ represents the index of $\mathcal{O}^{\alpha}_j$ listed in Table \ref{tbl-1} and since the operators are all those at the same site, the site index $j$ is omitted for simplicity.
	The spin-inversion parity index $p$ for ${\mathcal S}^n$ is also listed.}
	\centering
	\begin{tabular}{c l l l c l l l}\hline\hline
		\multicolumn{4}{c}{spin-1/2 KLM} & \multicolumn{4}{c}{spin-1 KLM} \\
		$n$ & \multicolumn{1}{c}{$\mathcal S^{n}_j$} & $\alpha$ & $p$ &
		$n$ & \multicolumn{1}{c}{$\mathcal S^{n}_j$} & $\alpha$ & $p$ \\ \hline
		a & $\ket{+}\bra{+}\hat{c}_{-}\hat{c}_{+}$ & 1 &$-$& a & $\ket{+}\bra{+}\hat{c}_{-}\hat{c}_{+}$ & 1 &$-$\\
		b & $\ket{+}\bra{-}\hat{c}_{-}\hat{c}_{+}$ & 2 &$+$& b & $\ket{+}\bra{0}\hat{c}_{-}\hat{c}_{+}$ & 2 &$-$\\
		c & $\ket{-}\bra{+}\hat{c}_{-}\hat{c}_{+}$ & 3 &$+$& c & $\ket{+}\bra{-}\hat{c}_{-}\hat{c}_{+}$ & 3 &$+$\\
		d & $\ket{-}\bra{-}\hat{c}_{-}\hat{c}_{+}$ & 4 &$-$& d & $\ket{0}\bra{+}\hat{c}_{-}\hat{c}_{+}$ & 4 &$-$\\
		1 & $\ket{+}\bra{+}\hat{c}_{+}$            & 1 &$+$& e & $\ket{0}\bra{-}\hat{c}_{-}\hat{c}_{+}$ & 5 &$+$\\
		2 & $\ket{+}\bra{+}\hat{c}_{-}$            & 1 &$-$& f & $\ket{0}\bra{0}\hat{c}_{-}\hat{c}_{+}$ & 6 &$-$\\
		3 & $\ket{+}\bra{-}\hat{c}_{+}$            & 2 &$-$& g & $\ket{-}\bra{+}\hat{c}_{-}\hat{c}_{+}$ & 7 &$+$\\
		4 & $\ket{+}\bra{-}\hat{c}_{-}$            & 2 &$+$& h & $\ket{-}\bra{0}\hat{c}_{-}\hat{c}_{+}$ & 8 &$+$\\
		5 & $\ket{-}\bra{+}\hat{c}_{+}$            & 3 &$-$& i & $\ket{-}\bra{-}\hat{c}_{-}\hat{c}_{+}$ & 9 &$-$\\
		6 & $\ket{-}\bra{+}\hat{c}_{-}$            & 3 &$+$& 1 & $\ket{+}\bra{+}\hat{c}_{+}$            & 1 &$+$\\
		7 & $\ket{-}\bra{-}\hat{c}_{+}$            & 4 &$+$& 2 & $\ket{+}\bra{+}\hat{c}_{-}$            & 1 &$-$\\
		8 & $\ket{-}\bra{-}\hat{c}_{-}$            & 4 &$-$& 3 & $\ket{+}\bra{0}\hat{c}_{+}$            & 2 &$+$\\
		  &                                        &   &   & 4 & $\ket{+}\bra{0}\hat{c}_{-}$            & 2 &$-$\\
		  &                                        &   &   & 5 & $\ket{+}\bra{-}\hat{c}_{+}$            & 3 &$-$\\
		  &                                        &   &   & 6 & $\ket{+}\bra{-}\hat{c}_{-}$            & 3 &$+$\\
		  &                                        &   &   & 7 & $\ket{0}\bra{+}\hat{c}_{+}$            & 4 &$+$\\
		  &                                        &   &   & 8 & $\ket{0}\bra{+}\hat{c}_{-}$            & 4 &$-$\\
		  &                                        &   &   & 9 & $\ket{0}\bra{0}\hat{c}_{+}$            & 5 &$+$\\
		  &                                        &   &   & 10 & $\ket{0}\bra{0}\hat{c}_{-}$           & 5 &$-$\\
		  &                                        &   &   & 11 & $\ket{0}\bra{-}\hat{c}_{+}$           & 6 &$-$\\
		  &                                        &   &   & 12 & $\ket{0}\bra{-}\hat{c}_{-}$           & 6 &$+$\\
		  &                                        &   &   & 13 & $\ket{-}\bra{+}\hat{c}_{+}$           & 7 &$-$\\
		  &                                        &   &   & 14 & $\ket{-}\bra{+}\hat{c}_{-}$           & 7 &$+$\\
		  &                                        &   &   & 15 & $\ket{-}\bra{0}\hat{c}_{+}$           & 8 &$-$\\
		  &                                        &   &   & 16 & $\ket{-}\bra{0}\hat{c}_{-}$           & 8 &$+$\\
		  &                                        &   &   & 17 & $\ket{-}\bra{-}\hat{c}_{+}$           & 9 &$+$\\
		  &                                        &   &   & 18 & $\ket{-}\bra{-}\hat{c}_{-}$           & 9 &$-$\\ \hline \hline
	\end{tabular}
	\label{list_coop}
\end{table}

\bibliography{ref}

\begin{thebibliography}{10}

\bibitem{UGe2}
S.~S. Saxena, P.~Agarwal, K.~Ahilan, F.~M. Grosche, R.~K.~W. Haselwimmer, M.~J.
  Steiner, E.~Pugh, I.~R. Walker, S.~R. Julian, P.~Monthoux, G.~G. Lonzarich,
  A.~Huxley, I.~Sheikin, D.~Braithwaite, and J.~Flouquet, Nature {\bfseries
  406},  587 (2000).

\bibitem{PRB.63.144519}
A.~Huxley, I.~Sheikin, E.~Ressouche, N.~Kernavanois, D.~Braithwaite,
  R.~Calemczuk, and J.~Flouquet, Phys. Rev. B {\bfseries 63},  144519 (2001).

\bibitem{URhGe}
D.~Aoki, A.~Huxley, E.~Ressouche, D.~Braithwaite, J.~Flouquet, J.-P. Brison,
  E.~Lhotel, and C.~Paulsen, Nature {\bfseries 413},  613 (2001).

\bibitem{UCoGe}
N.~T. Huy, A.~Gasparini, D.~E. de~Nijs, Y.~Huang, J.~C.~P. Klaasse,
  T.~Gortenmulder, A.~de~Visser, A.~Hamann, T.~G\"orlach, and H.~v.
  L\"ohneysen, Phys. Rev. Lett. {\bfseries 99},  067006 (2007).

\bibitem{JPSJ.73.3129}
T.~Akazawa, H.~Hidaka, H.~Kotegawa, T.~C.~Kobayashi, T.~Fujiwara, E.~Yamamoto,
  Y.~Haga, R.~Settai, and Y.~\~{O}nuki, J. Phys. Soc. Jpn. {\bfseries 73},
  3129 (2004).

\bibitem{Science.365.684}
S.~Ran, C.~Eckberg, Q.-P. Ding, Y.~Furukawa, T.~Metz, S.~R. Saha, I.-L. Liu,
  M.~Zic, H.~Kim, J.~Paglione, and N.~P. Butch, Science {\bfseries 365},  684
  (2019).

\bibitem{JPSJ.88.043702}
D.~Aoki, A.~Nakamura, F.~Honda, D.~Li, Y.~Homma, Y.~Shimizu, Y.~J. Sato,
  G.~Knebel, J.-P. Brison, A.~Pourret, D.~Braithwaite, G.~Lapertot, Q.~Niu,
  M.~Vališka, H.~Harima, and J.~Flouquet, J. Phys. Soc. Jpn. {\bfseries 88},
  043702 (2019).

\bibitem{Rev_Exp3}
D.~Aoki, K.~Ishida, and J.~Flouquet, J. Phys. Soc. Jpn. {\bfseries 88},  022001
  (2019).

\bibitem{PS75.546}
T.~Sakon, S.~Saito, K.~Koyama, S.~Awaji, I.~Sato, T.~Nojima, K.~Watanabe, and
  N.~K. Sato, Physica Scripta {\bfseries 75},  546 (2007).

\bibitem{PRL.100.077002}
N.~T. Huy, D.~E. de~Nijs, Y.~K. Huang, and A.~de~Visser, Phys. Rev. Lett.
  {\bfseries 100},  077002 (2008).

\bibitem{PRB.83.195107}
F.~Hardy, D.~Aoki, C.~Meingast, P.~Schweiss, P.~Burger, H.~v.~L\"ohneysen, and
  J.~Flouquet, Phys. Rev. B {\bfseries 83},  195107 (2011).

\bibitem{URhGe_RSC}
F.~L{\'e}vy, I.~Sheikin, B.~Grenier, and A.~D. Huxley, Science {\bfseries 309},
   1343 (2005).

\bibitem{PRL.114.216401}
Y.~Tokunaga, D.~Aoki, H.~Mayaffre, S.~Kr\"amer, M.-H. Julien, C.~Berthier,
  M.~Horvati\ifmmode~\acute{c}\else \'{c}\fi{}, H.~Sakai, S.~Kambe, and
  S.~Araki, Phys. Rev. Lett. {\bfseries 114},  216401 (2015).

\bibitem{PRB.93.201112}
Y.~Tokunaga, D.~Aoki, H.~Mayaffre, S.~Kr\"amer, M.-H. Julien, C.~Berthier,
  M.~Horvati\ifmmode~\acute{c}\else \'{c}\fi{}, H.~Sakai, T.~Hattori, S.~Kambe,
  and S.~Araki, Phys. Rev. B {\bfseries 93},  201112 (2016).

\bibitem{PRL.105.206403}
Y.~Ihara, T.~Hattori, K.~Ishida, Y.~Nakai, E.~Osaki, K.~Deguchi, N.~K. Sato,
  and I.~Satoh, Phys. Rev. Lett. {\bfseries 105},  206403 (2010).

\bibitem{PRL.107.187202}
C.~Stock, D.~A. Sokolov, P.~Bourges, P.~H. Tobash, K.~Gofryk, F.~Ronning, E.~D.
  Bauer, K.~C. Rule, and A.~D. Huxley, Phys. Rev. Lett. {\bfseries 107},
  187202 (2011).

\bibitem{PRL.108.066403}
T.~Hattori, Y.~Ihara, Y.~Nakai, K.~Ishida, Y.~Tada, S.~Fujimoto, N.~Kawakami,
  E.~Osaki, K.~Deguchi, N.~K. Sato, and I.~Satoh, Phys. Rev. Lett. {\bfseries
  108},  066403 (2012).

\bibitem{JPSJ.78.113709}
D.~Aoki, T.~D.~Matsuda, V.~Taufour, E.~Hassinger, G.~Knebel, and J.~Flouquet,
  J. Phys. Soc. Jpn. {\bfseries 78},  113709 (2009).

\bibitem{PhysRevB.22.3173}
D.~Fay and J.~Appel, Phys. Rev. B {\bfseries 22},  3173 (1980).

\bibitem{PhysRevB.67.024515}
T.~R. Kirkpatrick and D.~Belitz, Phys. Rev. B {\bfseries 67},  024515 (2003).

\bibitem{PhysRevB.66.134504}
V.~P. Mineev, Phys. Rev. B {\bfseries 66},  134504 (2002).

\bibitem{Rev_Theo1}
V.~P. Mineev, Phys.-Uspekhi {\bfseries 60},  121 (2017).

\bibitem{Rev_Theo2}
V.~P. Mineev, Low Temp. Phys. {\bfseries 44},  510 (2018).

\bibitem{PRL.90.167005}
K.~G. Sandeman, G.~G. Lonzarich, and A.~J. Schofield, Phys. Rev. Lett.
  {\bfseries 90},  167005 (2003).

\bibitem{PRL.94.097003}
A.~H. Nevidomskyy, Phys. Rev. Lett. {\bfseries 94},  097003 (2005).

\bibitem{PRB.77.184511}
J.~Linder, I.~B. Sperstad, A.~H. Nevidomskyy, M.~Cuoco, and A.~Sudb\o{}, Phys.
  Rev. B {\bfseries 77},  184511 (2008).

\bibitem{PRB.79.064501}
D.~V. Shopova and D.~I. Uzunov, Phys. Rev. B {\bfseries 79},  064501 (2009).

\bibitem{PRB.87.064501}
K.~Hattori and H.~Tsunetsugu, Phys. Rev. B {\bfseries 87},  064501 (2013).

\bibitem{JPCS.449.0120029}
Y.~Tada, S.~Fujimoto, N.~Kawakami, T.~Hattori, Y.~Ihara, K.~Ishida, K.~Deguchi,
  N.~K. Sato, and I.~Satoh, J. Phys. Conf. Ser. {\bfseries 449},  012029
  (2013).

\bibitem{PRB.93.174512}
Y.~Tada, S.~Takayoshi, and S.~Fujimoto, Phys. Rev. B {\bfseries 93},  174512
  (2016).

\bibitem{PRL.121.097001}
Y.~Sherkunov, A.~V. Chubukov, and J.~J. Betouras, Phys. Rev. Lett. {\bfseries
  121},  097001 (2018).

\bibitem{JPSJ.88.024707}
K.~Suzuki and K.~Hattori, J. Phys. Soc. Jpn. {\bfseries 88},  024707 (2019).

\bibitem{PRL.69.2863}
S.~R. White, Phys. Rev. Lett. {\bfseries 69},  2863 (1992).

\bibitem{PRB.48.10345}
S.~R. White, Phys. Rev. B {\bfseries 48},  10345 (1993).

\bibitem{PRL.105.036403}
Y.~Motome, K.~Nakamikawa, Y.~Yamaji, and M.~Udagawa, Phys. Rev. Lett.
  {\bfseries 105},  036403 (2010).

\bibitem{Yamamoto15704}
S.~J. Yamamoto and Q.~Si, Proc. Natl. Acad. Sci. U.S.A. {\bfseries 107},  15704
  (2010).

\bibitem{PRB.97.245119}
W.~Zhu and J.-X. Zhu, Phys. Rev. B {\bfseries 97},  245119 (2018).

\bibitem{PRB.75.165110}
S.~Saremi and P.~A. Lee, Phys. Rev. B {\bfseries 75},  165110 (2007).

\bibitem{PhysRevB.90.045125}
X.~Montiel, S.~Burdin, C.~P\'epin, and A.~Ferraz, Phys. Rev. B {\bfseries 90},
  045125 (2014).

\bibitem{S.R.7.11924}
N.~Xie, D.~Hu, and Y.-f. Yang, Sci. Rep. {\bfseries 7},  11924 (2017).

\bibitem{PRL.119.247203}
A.~M. Tsvelik and O.~M. Yevtushenko, Phys. Rev. Lett. {\bfseries 119},  247203
  (2017).

\bibitem{Bodensiek_2010}
O.~Bodensiek, T.~Pruschke, and R.~{\v{Z}}itko, J. Phys. Conf. Ser. {\bfseries
  200},  012162 (2010).

\bibitem{PRL.110.146406}
O.~Bodensiek, R.~\ifmmode~\check{Z}\else \v{Z}\fi{}itko, M.~Vojta, M.~Jarrell,
  and T.~Pruschke, Phys. Rev. Lett. {\bfseries 110},  146406 (2013).

\bibitem{PRL.115.036404}
J.~Otsuki, Phys. Rev. Lett. {\bfseries 115},  036404 (2015).

\bibitem{PRL.112.167204}
S.~Hoshino and Y.~Kuramoto, Phys. Rev. Lett. {\bfseries 112},  167204 (2014).

\bibitem{PhysRevB.79.220513}
P.~R. Bertussi, A.~L. Malvezzi, T.~Paiva, and R.~R. dos Santos, Phys. Rev. B
  {\bfseries 79},  220513 (2009).

\bibitem{JPSJ.75.043710}
S.~Watanabe, M.~Imada, and K.~Miyake, J. Phys. Soc. Jpn. {\bfseries 75},
  043710 (2006).

\bibitem{PRB.97.224519}
E.~K\c{a}dzielawa-Major, M.~Fidrysiak, P.~Kubiczek, and J.~Spa{\l}ek, Phys.
  Rev. B {\bfseries 97},  224519 (2018).

\bibitem{PRB.99.205106}
M.~Fidrysiak, D.~Goc-Jag{\l}o, E.~K\c{a}dzielawa-Major, P.~Kubiczek, and
  J.~Spa{\l}ek, Phys. Rev. B {\bfseries 99},  205106 (2019).

\bibitem{RMP.69.809}
H.~Tsunetsugu, M.~Sigrist, and K.~Ueda, Rev. Mod. Phys. {\bfseries 69},  809
  (1997).

\bibitem{PRL.77.3633}
S.~R. White, Phys. Rev. Lett. {\bfseries 77},  3633 (1996).

\bibitem{RKofRKKY}
M.~A. Ruderman and C.~Kittel, Phys. Rev. {\bfseries 96},  99 (1954).

\bibitem{KofRKKY}
T.~Kasuya, Prog. Theor. Phys. {\bfseries 16},  45 (1956).

\bibitem{YofRKKY}
K.~Yosida, Phys. Rev. {\bfseries 106},  893 (1957).

\bibitem{PRL.108.086402}
R.~Peters, N.~Kawakami, and T.~Pruschke, Phys. Rev. Lett. {\bfseries 108},
  086402 (2012).

\bibitem{PRB.86.165107}
R.~Peters and N.~Kawakami, Phys. Rev. B {\bfseries 86},  165107 (2012).

\bibitem{JPSJ.69.2947}
S.~Watanabe, J. Phys. Soc. Jpn. {\bfseries 69},  2947 (2000).

\bibitem{JPSCP.3.011032}
R.~Ishiyama and N.~Shibata, JPS Conf. Proc. {\bfseries 3},  011032 (2014).

\bibitem{PRB.46.13838}
M.~Sigrist, H.~Tsunetsugu, K.~Ueda, and T.~M. Rice, Phys. Rev. B {\bfseries
  46},  13838 (1992).

\bibitem{PRB.47.2886}
M.~Troyer and D.~W\"urtz, Phys. Rev. B {\bfseries 47},  2886 (1993).

\bibitem{PRB.47.8345}
H.~Tsunetsugu, M.~Sigrist, and K.~Ueda, Phys. Rev. B {\bfseries 47},  8345
  (1993).

\bibitem{PRL.78.2180}
G.~Honner and M.~Gul\'acsi, Phys. Rev. Lett. {\bfseries 78},  2180 (1997).

\bibitem{PRB.58.2662}
G.~Honner and M.~Gul\'acsi, Phys. Rev. B {\bfseries 58},  2662 (1998).

\bibitem{RMP.63.239}
M.~Sigrist and K.~Ueda, Rev. Mod. Phys. {\bfseries 63},  239 (1991).

\bibitem{PRB.97.134512}
S.~Sumita and Y.~Yanase, Phys. Rev. B {\bfseries 97},  134512 (2018).

\bibitem{PRB.94.174513}
T.~Nomoto, K.~Hattori, and H.~Ikeda, Phys. Rev. B {\bfseries 94},  174513
  (2016).

\bibitem{PRB.90.165114}
K.~Shiozaki and M.~Sato, Phys. Rev. B {\bfseries 90},  165114 (2014).

\bibitem{arXiv:1909.09634}
S.~Ono, H.~C. Po, and H.~Watanabe, arXiv:1909.09634.

\bibitem{JETPL.20.287}
V.~L. Berezinskii, JETP. Lett. {\bfseries 20},  287 (1974).

\bibitem{arXiv1709.03986}
J.~Linder and A.~V. Balatsky, arXiv1709.03986.

\bibitem{PRB.48.7445}
A.~V. Balatsky and J.~Bon\v{c}a, Phys. Rev. B {\bfseries 48},  7445 (1993).

\bibitem{PRB.52.1271}
E.~Abrahams, A.~Balatsky, D.~J. Scalapino, and J.~R. Schrieffer, Phys. Rev. B
  {\bfseries 52},  1271 (1995).

\bibitem{Dahal_2009}
H.~P. Dahal, E.~Abrahams, D.~Mozyrsky, Y.~Tanaka, and A.~V. Balatsky, New J.
  Phys. {\bfseries 11},  065005 (2009).

\bibitem{PhysRevLett.77.1342}
O.~Zachar, S.~A. Kivelson, and V.~J. Emery, Phys. Rev. Lett. {\bfseries 77},
  1342 (1996).

\bibitem{PRB.100.094532}
S.~Iimura, M.~Hirayama, and S.~Hoshino, Phys. Rev. B {\bfseries 100},  094532
  (2019).

\bibitem{PhysRevB.83.205113}
A.~Moreno, A.~Muramatsu, and S.~R. Manmana, Phys. Rev. B {\bfseries 83},
  205113 (2011).

\bibitem{PhysRevB.98.140505}
H.-C. Jiang, Z.-Y. Weng, and S.~A. Kivelson, Phys. Rev. B {\bfseries 98},
  140505 (2018).

\bibitem{Jiang1424}
H.-C. Jiang and T.~P. Devereaux, Science {\bfseries 365},  1424 (2019).

\bibitem{PRL.9.290}
V.~Jaccarino and M.~Peter, Phys. Rev. Lett. {\bfseries 9},  290 (1962).

\end{thebibliography}

\end{document}